\documentclass[pra,twocolumn,superscriptaddress,showpacs,nofootinbibfloatfix,amsmath,amsfonts,amssymb]{revtex4-1}%
\usepackage[T1]{fontenc}
\usepackage[latin9]{inputenc}
\usepackage{array}
\usepackage{multirow}
\usepackage{amsmath}
\usepackage{amssymb}
\usepackage{graphicx}

\makeatletter

\pdfpageheight\paperheight
\pdfpagewidth\paperwidth

\providecommand{\tabularnewline}{\\}

\makeatother

\usepackage{amsmath,amsfonts,amssymb,color}
\usepackage{amsthm}
\usepackage{leftidx}
\usepackage{graphicx}
\usepackage{xcolor}
\usepackage{dcolumn}
\usepackage{bm}
\usepackage{epstopdf}
\usepackage{epsfig}
\usepackage{environ}
\usepackage{pdfcomment}

\usepackage{float}
\usepackage[T1]{fontenc}
\usepackage[latin9]{inputenc}
\usepackage{setspace}
\usepackage{esint}

\begin{document}
\preprint{APS/123-QED}
\title{Floquet engineering of topological localization transitions and \protect\\ mobility edges in one-dimensional non-Hermitian quasicrystals}
\author{Longwen Zhou}
\email{zhoulw13@u.nus.edu}

\affiliation{College of Physics and Optoelectronic Engineering,
Ocean University of China, Qingdao, China 266100}
\date{\today}
\begin{abstract}
Time-periodic driving fields could endow a system with peculiar topological
and transport features. In this work, we find dynamically controlled
localization transitions and mobility edges in non-Hermitian quasicrystals via shaking
the lattice periodically. The driving force dresses the
hopping amplitudes between lattice sites, yielding alternate transitions
between localized, mobility edge and extended non-Hermitian quasicrystalline phases. We apply
our Floquet engineering approach to five representative models of
non-Hermitian quasicrystals, obtain the conditions of ``photon-assisted''
localization transitions and mobility edges, and find the expressions of Lyapunov exponents for some models. We further introduce topological winding
numbers of Floquet quasienergies to distinguish non-Hermitian quasicrystalline
phases with different localization nature. Our discovery thus extend
the study of quasicrystals to non-Hermitian Floquet systems, and provide
an efficient way of modulating the topological and transport properties
of these unique phases.
\end{abstract}
\maketitle

\section{Introduction\label{sec:Int}}

Floquet engineering has enabled the realization of rich dynamical,
topological and transport phenomena in a broad range of physical settings~\cite{FloExp0,FloExp1,FloExp2,FloExp3,FloExp4,FloExp5,FloExp6,FloExp7,FloExp8,FloExp9,FloExp10,FloExp11,FloExp12,FloExp13,FloExp14,FloExp15,FloExp16}.
Some notable examples include the dynamical localization, Floquet
topological phase and discrete time crystal (see~\cite{DLRev1,DLRev2,DLRev3,DLRev5,FTPRev1,FTPRev2,FTPRev3,FTPRev4,FTPRev5,DTCRev1,DTCRev2,DTCRev3,DTCRev4}
for reviews and references therein). These discoveries have broadened
the classification of phases of matter in nonequilibrium situations~\cite{FTPClass1,FTPClass2,FTPClass3}, and may lead to applications
in emerging technologies like ultrafast electronics~\cite{FTPRev1}
and topological quantum computing~\cite{FTQC1,FTQC2,FTQC3}.

Non-Hermitian topological matter has been investigated intensively in the past few years~\cite{NHRev1,NHRev2,NHRev3,NHTP1,NHTP2,NHTP3}. Recently, the Floquet approach was also applied to engineer topological
phases in non-Hermitian systems. Notably, it was found that the interplay
between time-periodic driving fields and gain and loss or nonreciprocal
effects could yield topological phases that are unique to non-Hermitian
Floquet systems~\cite{NHFTP1,NHFTP2,NHFTP3,NHFTP4,NHFTP5,NHFTP6,NHFTP7,NHFTP8}. These intriguing phases are characterized
by large integer or half-integer winding numbers and degenerate Floquet
edge or corner states with real quasienergies~\cite{NHFTP1,NHFTP2,NHFTP4,NHFTP5,NHFTP6,NHFTP7,NHFTP8}. Their topological signatures
may further be extracted experimentally from the dynamical spin textures
and quantized displacements of wavepackets~\cite{NHFTP3}. Till now, various kinds
of non-Hermitian Floquet topological insulators, superconductors and
semimetals with exceptional topological properties have been found~\cite{NHFTP1,NHFTP2,NHFTP3,NHFTP4,NHFTP5,NHFTP6,NHFTP7,NHFTP8,NHFTP9,NHFTP10,NHFTP11,NHFTP12,NHFTP13,NHFTP14,NHFTP15,NHFTP16,NHFTP17,NHFTP18,NHFTP19,NHFTP20,NHFTP21,NHFTP22,NHFTP23,NHFTP24,NHFTP25,NHFTP26,NHFTP27,NHFTP28,NHFTP29,NHFTP30}. Meanwhile, much less is known when non-Hermitian
Floquet systems are subject to more complicated correlation effects,
such as disorder, nonlinearity and many-body interactions. 

In this work, we employ the idea of Floquet engineering to induce and control the
spectrum, localization, mobility edges and topological transitions in non-Hermitian
quasicrystals. A quasicrystal is a phase of matter with long-range order in the absence
of spatial periodicity. It can be realized in a lattice with diagonal
or off-diagonal spatially quasiperiodic modulations. In Sec.~\ref{sec:Met}, we outline the method of adjusting
hopping amplitudes in a tight-binding lattice by applying high-frequency periodic
driving forces, which follows the idea of engineering dynamical localization
in ultracold atoms~\cite{FloExp0,FloExp1,FloExp2}. In Sec.~\ref{sec:Res}, we introduce five prototypical
models of non-Hermitian quasicrystal, which can be viewed as non-Hermitian
extensions of the Aubry-André-Harper (AAH) model and the Maryland model. We further apply high-frequency
shaking forces to these models, and obtain driving-induced transitions
between non-Hermitian Floquet quasicrystalline phases with distinct
transport and topological features. Different types of non-Hermitian
quasicrystal phases are also found to emerge alternately with the
increase of the ratio between the amplitude and frequency of the driving
field. In Sec.~\ref{sec:Sum}, we summarize our results and discuss future
perspectives. 

\section{Method\label{sec:Met}}
\begin{figure}[b]
	\centering{}\includegraphics[scale=0.55]{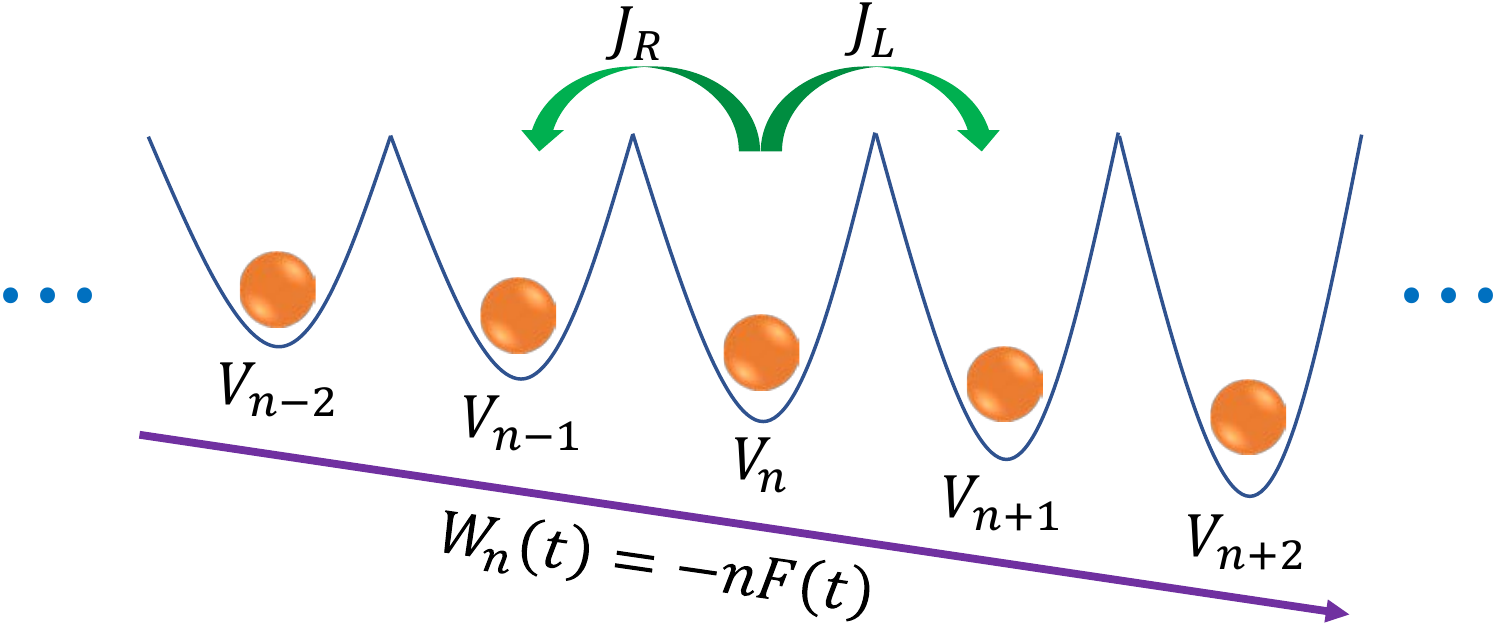}\caption{A schematic illustration of the driven non-Hermitian superlattice. $J_{L}$
		($J_{R}$) denotes the hopping amplitude from left (right) to right
		(left) lattice sites. $V_{n}$ denotes the static potential on the $n$th
		lattice site, which introduces quasiperiodicity into the system. $W_{n}(t)$
		is the driving potential, which is proportional to the position
		index $n$. The driving force $F(t)$ oscillates periodically in time.
		The system is made non-Hermitian if $J_{L}\protect\neq J_{R}^{*}$
		or $V_{n}\protect\neq V_{n}^{*}$, corresponding to the case with
		nonreciprocal hopping or onsite gain and loss.\label{fig:Sketch}}
\end{figure}

In this section, we introduce our scheme of Floquet engineering to
control the phase transitions in non-Hermitian quasicrystals, which is based
on the idea of realizing dynamical localization in Bose-Einstein condensates~\cite{DLRev2}. 
A schematic illustration of our system and approach is
shown in Fig.~\ref{fig:Sketch}. The Hamiltonian of the system describes
a tight-binding superlattice with onsite quasiperiodic potential and
driving field, i.e., 
\begin{equation}
\hat{H}(t)=\sum_{\langle n,n'\rangle}(J_{R}\hat{c}_{n}^{\dagger}\hat{c}_{n'}+J_{L}\hat{c}_{n'}^{\dagger}\hat{c}_{n})+\sum_{n}[V_{n}+W_{n}(t)]\hat{c}_{n}^{\dagger}\hat{c}_{n},\label{eq:Ht}
\end{equation}
where $n$ is the lattice site index, and $\langle n,n'\rangle$ includes site indices with $n'>n$. $\hat{c}_{n}^{\dagger}$
and $\hat{c}_{n}$ denote the creation and annihilation operators
of a particle (either boson or fermion) on the $n$th lattice site. $J_{L}$ ($J_{R}$) represents
the hopping amplitude of the particle from lattice site $n$~($n'$)
to $n'$~($n$). A quasicrystal model is realized by setting the superlattice
potential $V_{n}$ to a quasiperiodic function of the lattice index $n$.
A non-Hermitian quasicrystal is further obtained by setting $J_{L}\neq J_{R}^{*}$
(for nonreciprocal hopping) or $V_{n}\neq V_{n}^{*}$ (for gain and
loss of particles). The coupling between the driving field and the
particle takes the form of $W_{n}(t)=-nF(t)$, where the driving force $F(t)=F(t+T)$
varies periodically in time with the driving period $T$ and driving
frequency $\omega=2\pi/T$. All system parameters have been properly rescaled
and given in dimensionless units. We will also set the Planck
constant $\hbar=1$ throughout the discussion.

In the frame comoving with the lattice,
the dynamics of the system is governed by the time-dependent
Schr\"odinger equation
\begin{equation}
i\frac{d}{dt}|\psi(t)\rangle=\hat{H}(t)|\psi(t)\rangle.\label{eq:SeqLab}
\end{equation}
The driving potential in $\hat{H}(t)$ can be transformed to an oscillating
phase factor in the hopping amplitudes $J_{L,R}$~\cite{DLRev2}. To do so, we consider
the following unitary rotation
\begin{equation}
\hat{R}(t)=e^{i\sum_{n}f_{n}(t)\hat{c}_{n}^{\dagger}\hat{c}_{n}},\label{eq:Rot}
\end{equation}
where the time-dependent amplitude 
\begin{alignat}{1}
f_{n}(t)= & -\int_{t_{0}}^{t}W_{n}(t')dt'-n\overline{f}\label{eq:ft}\\
= & n\left(\int_{t_{0}}^{t}F(t')dt'-\overline{f}\right)\equiv nf(t).\nonumber 
\end{alignat}
Here $\overline{f}$ is chosen to ensure that the one-period integral
of amplitude $\int_{0}^{T}f_n(t)dt=0$. Writing the time-evolving
state as $|\psi(t)\rangle=\hat{R}(t)|\psi'(t)\rangle$, we obtain
the Schr\"odinger equation for the state $|\psi'(t)\rangle$ in the
rotating frame as
\begin{equation}
i\frac{d}{dt}|\psi'(t)\rangle=\hat{H}_{{\rm r}}(t)|\psi'(t)\rangle,\label{eq:SeqRot}
\end{equation}
where the Hamiltonian $\hat{H}_{{\rm r}}(t)$ in the rotating frame reads
\begin{equation}
\hat{H}_{{\rm r}}(t)=\hat{R}^{\dagger}(t)\hat{H}(t)\hat{R}(t)-i\hat{R}^{\dagger}(t)\frac{d\hat{R}(t)}{dt}.\label{eq:HR}
\end{equation}
Plugging Eqs.~(\ref{eq:Rot}) and (\ref{eq:ft}) into Eq.~(\ref{eq:HR}),
we find that $\hat{R}(t)$ commutes with the onsite terms $\sum_{n}[V_{n}+W_{n}(t)]\hat{c}_{n}^{\dagger}\hat{c}_{n}$
in $\hat{H}(t)$ and 
\begin{equation}
i\hat{R}^{\dagger}(t)\frac{d\hat{R}(t)}{dt}=\sum_{n}W_{n}(t)\hat{c}_{n}^{\dagger}\hat{c}_{n},\label{eq:EqRot}
\end{equation}
which cancels the driving potential $\sum_{n}W_{n}(t)\hat{c}_{n}^{\dagger}\hat{c}_{n}$,
leaving $\sum_{n}V_{n}\hat{c}_{n}^{\dagger}\hat{c}_{n}$ as the only
onsite term in $\hat{H}_{{\rm r}}(t)$. Furthermore, the hopping terms
in $\hat{H}(t)$ under the rotation $\hat{R}(t)$ becomes
\begin{alignat}{1}
\hat{R}^{\dagger}(t)\hat{c}_{n}^{\dagger}\hat{c}_{n'}\hat{R}(t)= & e^{-inf(t)\hat{c}_{n}^{\dagger}\hat{c}_{n}}\hat{c}_{n}^{\dagger}e^{inf(t)\hat{c}_{n}^{\dagger}\hat{c}_{n}}\nonumber \\
\times & e^{-in'f(t)\hat{c}_{n'}^{\dagger}\hat{c}_{n'}}\hat{c}_{n'}e^{in'f(t)\hat{c}_{n'}^{\dagger}\hat{c}_{n'}}\nonumber \\
= & e^{i(n'-n)f(t)}\hat{c}_{n}^{\dagger}\hat{c}_{n'}.\label{eq:EqHop}
\end{alignat}
Note that this result holds for both bosons and fermions. Therefore, the Hamiltonian in the rotating frame is
\begin{alignat}{1}
\hat{H}_{{\rm r}}(t)= & \sum_{\langle n,n'\rangle}J_{L}e^{i(n-n')f(t)}\hat{c}_{n'}^{\dagger}\hat{c}_{n}\nonumber \\
+ & \sum_{\langle n,n'\rangle}J_{R}e^{i(n'-n)f(t)}\hat{c}_{n}^{\dagger}\hat{c}_{n'}+\sum_{n}V_{n}\hat{c}_{n}^{\dagger}\hat{c}_{n},\label{eq:HR2}
\end{alignat}
where the oscillating phase factor $f(t)=\int_{t_{0}}^{t}F(t')dt'-\overline{f}$.
When $f(t)$ varies quickly in time, so that $\hbar\omega\gg |J_{R,L}|$ and $|V_n-V_{n'}|$ for $|n-n'|=1$,
its effect on $\hat{H}_{{\rm r}}(t)$
can be approximated by the average over a complete driving period
$T$~\cite{DLRev2}. This corresponds to taking the zeroth order term in the high-frequency
expansion of the Floquet Hamiltonian~\cite{HFA1,HFA2,HFA3}. In this limit, the micromotion
operator is approximated by the rotating frame transformation $\hat{R}(t)$,
and the resulting Floquet effective Hamiltonian reads
\begin{alignat}{1}
\hat{H}_{{\rm F}} & \approx\frac{1}{T}\int_{0}^{T}dt\hat{H}_{{\rm r}}(t)\nonumber \\
 & =\sum_{\langle n,n'\rangle}\left(J_{L}^{{\rm eff}}\hat{c}_{n'}^{\dagger}\hat{c}_{n}+J_{R}^{{\rm eff}}\hat{c}_{n}^{\dagger}\hat{c}_{n'}\right)+\sum_{n}V_{n}\hat{c}_{n}^{\dagger}\hat{c}_{n},\label{eq:HFG}
\end{alignat}
where the ``photon-dressed'' hopping amplitudes $J_{L}^{{\rm eff}}$
and $J_{R}^{{\rm eff}}$ are given by
\begin{alignat}{1}
J_{L}^{{\rm eff}}= & \frac{1}{T}\int_{0}^{T}dtJ_{L}e^{i(n-n')f(t)},\label{eq:JLeff}\\
J_{R}^{{\rm eff}}= & \frac{1}{T}\int_{0}^{T}dtJ_{R}e^{i(n'-n)f(t)}.\label{eq:JReff}
\end{alignat}
The driving field thus introduces a control knob to the hopping amplitudes, and thus the transport property
of the non-Hermitian quasicrystal.

In experiments, one type of driving force that is relatively easy
to engineer has the form of a harmonic function~\cite{FloExp0,DLRev2}, which can be
introduced by shaking the lattice back and forth periodically~\cite{HFA4,HFA5,HFA6} and described
as $F(t)=K\cos(\omega t)$, where $K$ is the driving amplitude and
$\omega$ is the driving frequency of the force. The factor $f(t)$ in the Peierls phase for such
a harmonic drive reads $f(t)=K\sin(\omega t)/\omega$ according to Eq.~(\ref{eq:ft}). Working
out the the integrals in Eqs.~(\ref{eq:JLeff}) and (\ref{eq:JReff})
with the help of the Bessel function expansion $e^{ix\sin(y)}=\sum_{l=-\infty}^{\infty}{\cal J}_{l}(x)e^{ily}$,
we find $J_{L,R}^{{\rm eff}}=J_{L,R}{\cal J}_{0}\left((n'-n)K/\omega\right)$, which is
derived previously in the study of dynamical localization~\cite{HFA4}. 
The Floquet effective Hamiltonian of the system under the high-frequency
harmonic forcing now takes the form
\begin{alignat}{1}
\hat{H}_{{\rm F}}= & \sum_{\langle n,n'\rangle}J_{L}{\cal J}_{0}\left((n'-n)\frac{K}{\omega}\right)\hat{c}_{n'}^{\dagger}\hat{c}_{n}\label{eq:HF0}\\
+ & \sum_{\langle n,n'\rangle}J_{R}{\cal J}_{0}\left((n'-n)\frac{K}{\omega}\right)\hat{c}_{n}^{\dagger}\hat{c}_{n'}+\sum_{n}V_{n}\hat{c}_{n}^{\dagger}\hat{c}_{n}.\nonumber 
\end{alignat}
Since both the $J_{L,R}$ and the Bessel function of first kind ${\cal J}_{0}\left((n'-n)K/\omega\right)$
decays globaly with the increase of hopping distance $|n'-n|$, we
may reserve the hopping terms up to nearest-neighbor sites, yielding
the following Floquet effective Hamiltonian
\begin{equation}
\hat{H}_{{\rm F}}=\sum_{n}\left[{\cal J}_{0}\left(\frac{K}{\omega}\right)\left(J_{R}\hat{c}_{n}^{\dagger}\hat{c}_{n+1}+J_{L}\hat{c}_{n+1}^{\dagger}\hat{c}_{n}\right)+V_{n}\hat{c}_{n}^{\dagger}\hat{c}_{n}\right].\label{eq:HF}
\end{equation}
In the Hermitian limit, we observe that the hopping amplitudes $J_{L,R}$
are switched off at the zeros of ${\cal J}_{0}(K/\omega)$, which
leads to the well-known phenomena of dynamical localization for any
nonvanishing onsite potential $V_{n}$~\cite{FloExp0}. When the system becomes non-Hermitian
and $V_{n}$ is quasiperiodic, the competition between the energy scales of intersite
tunneling and onsite trapping could result in localization-delocalization
transitions caused by non-Hermitian effects. The presence of the driving
field further generates an interplay between the field parameters
$(K,\omega)$ and the non-Hermitian terms of the system, which is
expected to yield richer phase diagrams and dynamically controlled
transitions between different non-Hermitian quasicrystalline phases.
These judgments will be demonstrated by explicit examples in the
following section.

\section{Results\label{sec:Res}}
In this section, we apply the Floquet engineering approach discussed
in Sec.~\ref{sec:Met} to five representative models of non-Hermitian
quasicrystal, which can be viewed as non-Hermitian variants of
the AAH model~\cite{AAH1,AAH2,AAH3} and the Maryland model~\cite{MM1,MM2,MM3,MM5}. The explicit forms of these models
and the driving forces are introduced in Subsec.~\ref{subsec:NHFQC}.
Without the driving field, the conditions of localization transition
for these models have been derived in previous studies. We reveal
how these conditions are modified in the presence of high-frequency
lattice shaking forces, and obtain analytical expressions for their
Lyapunov exponents if possible, which are now functions of the driving amplitude
and frequency. Our theoretical results are further verified
by numerical calculations presented in Subsecs.~\ref{subsec:M1}--\ref{subsec:M5}.

\subsection{Quasicrystal models\label{subsec:NHFQC}}

\begin{table*}
	\begin{centering}
		\begin{tabular}{|c|c|c|c|c|}
			\hline 
			Model & Hopping & Onsite & Driving\tabularnewline
			index & amplitude & potential $V_{n}$ & force $W_{n}(t)$\tabularnewline
			\hline 
			\hline 
			M1 & \multirow{2}{*}{$J_{L}=J_{R}=J$} & $Ve^{-i2\pi\alpha n}$ & \multirow{5}{*}{$-nK\cos(\omega t)$}\tabularnewline
			\cline{1-1} \cline{3-3} 
			M2 &  & $V\cos(2\pi\alpha n+i\gamma)$ & \tabularnewline
			\cline{1-3}  
			M3 & $J_{L}=Je^{\gamma},\,J_{R}=Je^{-\gamma}$ & $V\cos(2\pi\alpha n)$ & \tabularnewline
			\cline{1-3}
			M4 & \multirow{2}{*}{$J_{L}=J_{R}=J$} & $V\tan(\pi\alpha n+i\gamma)$ & \tabularnewline
			\cline{1-1} \cline{3-3} 
			M5 &  & $V/\left(1-\eta e^{i2\pi\alpha n}\right)$ & \tabularnewline
			\hline 
		\end{tabular}
		\par\end{centering}
	\caption{Definitions of the five periodically forced non-Hermitian quasicrystal
		models, denoted by M1--M5. In the second column, $J_{L}$ and $J_{R}$ are left-to-right
		and right-to-left hopping amplitudes between nearest-neighbor lattice
		sites. $\gamma$ controls the asymmetry between the left and right
		hopping amplitudes in M3. In the third column, $V$ is the amplitude
		of the onsite potential, $i\gamma$ is an imaginary phase shift, $\eta$ controls the non-Hermiticity of M5 and
		$\alpha$ is an irrational number. In the fourth column, $n$ is the
		lattice site index, $K$ is the driving amplitude and $\omega$ is
		the driving frequency. All system parameters are set in dimensionless
		units.\label{tab:ModelDef}}
\end{table*}

The AAH model is one of the standard models in the study of localization-delocalization
transitions in one-dimensional (1D) quasiperiodic systems~\cite{AAH1,AAH2,AAH3}. In the
lattice representation, the Hamiltonian of the model takes the form
\begin{equation}
\hat{H}_{0}=\sum_{n}\left[J\left(\hat{c}_{n}^{\dagger}\hat{c}_{n+1}+{\rm h.c.}\right)+V\cos(2\pi\alpha n)\hat{c}_{n}^{\dagger}\hat{c}_{n}\right],\label{eq:HAAH}
\end{equation}
where $J$ is the hopping amplitude and $V$ is the amplitude of the
onsite potential. By setting $\alpha$ to be an irrational number,
the superlattice potential $V\cos(2\pi\alpha n)$ becomes quasiperiodic
in the lattice index $n$, and the system described by $\hat{H}_{0}$
forms a 1D quasicrystal. Thanks to the self-duality property of the
AAH quasicrystal, it has been shown that the eigenstates of $\hat{H}_{0}$
possess a delocalization-to-localization transition at $V=2J$ in the
thermodynamic limit. When $V<2J$, all eigenstates of $\hat{H}_{0}$
are delocalized and the system is in an extended phase. When $V>2J$,
the correlated disorder becomes
strong enough, such that all eigenstates of $\hat{H}_{0}$ are localized
and the system enters an insulating phase~\cite{AAH3}.

The Maryland model forms another paradigm in the study of Anderson localization and quantum chaos~\cite{MM1,MM2,MM3,MM5}. It is
an integrable model and can be mapped to a Floquet system. The Hamiltonian of the Maryland model takes the form
\begin{equation}
\hat{H}_{\rm M}=\sum_{n}\left[J\left(\hat{c}_{n}^{\dagger}\hat{c}_{n+1}+{\rm h.c.}\right)+V\tan(\pi\alpha n)\hat{c}_{n}^{\dagger}\hat{c}_{n}\right],\label{eq:HMM}
\end{equation}
where $n$ is the lattice index, $J$ is the hopping amplitude, $V$ is the amplitude of the onsite potential, and a quasicrystal is formed once $\alpha$ takes an irrational value. Since the potential $V_n$ depends on a tangent function, which is not bounded, all states of the system are localized and are found to have an energy-dependent Lyapunov exponent~(inverse localization length) $\lambda(E)={\rm arccosh}[(\sqrt{(2J+E)^2+V^2}+\sqrt{(2J-E)^2+V^2})/4|J|]$ \cite{MM1}. Besides, the Maryland model can also be employed to study the topoloigcal nature of integer quantum Hall effects~\cite{MM4}.

In recent years, the impact of non-Hermiticity on localization transitions
in AAH-type and Maryland-type quasicrystals have been explored~\cite{MM3,LonghiQC1,LonghiQC2,LonghiQC3,LonghiQC4,ChenQC1,ChenQC2,ChenQC3,ChenQC4,ChenQC5,ChenQC6,CaiQC1,NHQC1,NHQC2,NHQC3,NHQC4,NHQC5,NHQC6,NHQC7,NHQC8,NHQC9,NHQC10}. The non-Hermitian effects
are introduced by either setting the onsite quasiperiodic potential
to be non-Hermitian~\cite{MM3,LonghiQC1,LonghiQC2,LonghiQC3}, or making the hopping amplitudes between adjacent
lattice sites to be nonreciprocal~\cite{LonghiQC4,ChenQC1}. In these models, it was found
that the non-Hermitian terms could induce ${\cal PT}$-transitions
of the energy spectrum from real to complex (or the opposite), together with
localization-delocalization transitions of the eigenstates.
These transitions could further be characterized by the quantized change
of winding numbers of the spectrum around certain base points of the complex
energy plane~\cite{MM3,LonghiQC1,LonghiQC2,LonghiQC3,LonghiQC4,ChenQC1}. 

In the following subsections, we apply the Floquet engineering approach to induce
and control phase transitions in five non-Hermitian quasicrystal
models. The time-dependent Hamiltonians of these models take the general
form of Eq.~(\ref{eq:Ht}), with the explicit expressions of hopping
amplitudes, onsite potential and driving force given in Table~\ref{tab:ModelDef}.
We will refer to these five models as M1--M5 for simplicity.
In the high-frequency limit, the Floquet effective Hamiltonians of
these models then take the formalism of Eq.~(\ref{eq:HF}), where
the hopping amplitudes are controlled by the ratio of the driving amplitude
and driving frequency $K/\omega$ through the Bessel function ${\cal J}_{0}(K/\omega)$.
Since the localization-delocalization transitions in these models
are originated from the competition between the energy scales of hopping
$J$ and onsite potential $V$~\cite{AAH3}, the ``photon-dressed'' hopping amplitude
$J_{L,R}{\cal J}_{0}(K/\omega)$ provides a flexible knob to tune
the transitions between different non-Hermitian quasicrystalline phases of these systems,
as will be demonstrated below.
Note in passing that the models M1--M5 do not possess the translation symmetry. They also do not have the chiral symmetry due to the absence of sublattice or spin structures. Meanwhile, the effective Hamiltonians of M1--M5 all possess the $\cal{PT}$-symmetry, which is relevant for the characterization of their phase transitions.

\subsection{M1: Floquet spectrum, localization transition and topological invariant\label{subsec:M1}}

Following Table~\ref{tab:ModelDef}, the time-dependent Hamiltonian
of the quasicrystal M1 reads
\begin{alignat}{1}
\hat{H}_{1}(t)= & \sum_{n}J\left(\hat{c}_{n}^{\dagger}\hat{c}_{n+1}+{\rm h.c.}\right)\nonumber \\
+ & \sum_{n}\left[Ve^{-i2\pi\alpha n}-nK\cos(\omega t)\right]\hat{c}_{n}^{\dagger}\hat{c}_{n},\label{eq:H1t}
\end{alignat}
which is non-Hermitian due to the complex onsite potential $Ve^{-i2\pi\alpha n}$
and quasiperiodic when $\alpha$ is irrational. Following the procedure
of Sec.~\ref{sec:Met}, we find the Floquet effective Hamiltonian
of the system under high-frequency driving as
\begin{equation}
\hat{H}_{1{\rm F}}=\sum_{n}\left[J{\cal J}_{0}\left(\frac{K}{\omega}\right)\left(\hat{c}_{n}^{\dagger}\hat{c}_{n+1}+{\rm h.c.}\right)+Ve^{-i2\pi\alpha n}\hat{c}_{n}^{\dagger}\hat{c}_{n}\right].\label{eq:H1F}
\end{equation}
The hopping strength now depends on the ratio between the amplitude and
frequency of the driving field, and can thus be controlled dynamically.
Meanwhile, the effective Hamiltonian $\hat{H}_{1{\rm F}}$ possesses
the ${\cal PT}$-symmetry as $V_{n}=V_{-n}^{*}$, which means that
its Floquet spectrum could be real in certain parameter regions.

\begin{table*}
	\begin{centering}
		\begin{tabular}{|c|c|c|}
			\hline 
			Phase & Extended & Localized\tabularnewline
			\hline 
			\hline 
			Condition & $|V|<|J{\cal J}_{0}(K/\omega)|$ & $|V|>|J{\cal J}_{0}(K/\omega)|$\tabularnewline
			\hline 
			IPR & $\simeq0$ for all states & $>0$ for all states\tabularnewline
			\hline 
			Lyapunov exponent & \multicolumn{2}{c|}{$\lambda=\ln\left|\frac{V}{J{\cal J}_{0}(K/\omega)}\right|\begin{cases}
				<0 & {\rm Extended}\\
				>0 & {\rm Localized}
				\end{cases}$}\tabularnewline
			\hline 
			Quasienergy & $E(k)=2J{\cal J}_{0}(K/\omega)\cos k$ & $E(k)=2J{\cal J}_{0}(K/\omega)\cos(k-i\lambda)$\tabularnewline
			\hline 
			Winding number & \multicolumn{2}{c|}{$w=\int_{0}^{2\pi}\frac{d\theta}{2\pi i}\partial_{\theta}\ln\det[\hat{H}'_{1{\rm F}}(\theta)-E_{0}]=\begin{cases}
				0 & {\rm Extended}\\
				-1 & {\rm Localized}
				\end{cases}$}\tabularnewline
			\hline 
		\end{tabular}
		\par\end{centering}
	\caption{Summary of the results for the non-Hermitian Floquet quasicrystal M1.
		$J$ is the nearest-neighbor hopping amplitude, $V$ is the amplitude
		of onsite non-Hermitian quasiperiodic potential, $K$ is the driving
		amplitude, $\omega$ is the driving frequency, and ${\cal J}_{0}(K/\omega)$
		denotes the Bessel function of first kind. $\lambda$ denotes the Lyapunov exponent~\cite{Note1}. The parameter $k$ fills
		uniformly the range of $[0,2\pi)$. $\hat{H}'_{1{\rm F}}(\theta)$
		is obtained from $\hat{H}_{1{\rm F}}$ after taking the discrete Fourier
		transformation and introducing a phase twist $e^{-i\theta}$ to its
		top-right corner matrix element in momentum representation. $E_{0}$
		is the base quasienergy.\label{tab:ResM1}}
\end{table*}

The eigenvalue equation of $\hat{H}_{1{\rm F}}$ is $\hat{H}_{{\rm 1F}}|\psi\rangle=E|\psi\rangle$. Inserting the single-particle state $|\psi\rangle=\sum_{n}\psi_{n}\hat{c}_{n}^{\dagger}|0\rangle$ into the equation,
we obtain the eigenvalue equation in the lattice representation as
\begin{equation}
J{\cal J}_{0}\left(\frac{K}{\omega}\right)(\psi_{n+1}+\psi_{n-1})+Ve^{-i2\pi\alpha n}\psi_{n}=E\psi_{n}.\label{eq:H1E}
\end{equation}
Here $E\mod2\pi$ corresponds to the quasienergy of $\hat{H}_{1{\rm F}}$.
For a lattice of length $L$ and under the periodic boundary condition
(PBC), one can take a discrete Fourier transformation from lattice
to momentum representations by expanding the amplitude $\psi_{n}$
as
\begin{equation}
\psi_{n}=\frac{1}{\sqrt{L}}\sum_{m=1}^{L}\varphi_{m}e^{-i2\pi\alpha mn}.\label{eq:FFT}
\end{equation}
The transformed eigenvalue equation in momentum space then reads
\begin{equation}
V\varphi_{n-1}+2J{\cal J}_{0}\left(\frac{K}{\omega}\right)\cos(2\pi\alpha n)\varphi_{n}=E\varphi_{n}.\label{eq:H1M}
\end{equation}
Following the method outlined in Ref.~\cite{LonghiQC2}, it can be readily shown
that when $\alpha$ takes irrational values, the non-Hermitian Floquet
quasicrystal M1 possesses a ${\cal PT}$-transition and a localization-delocalization
transition under the condition 
\begin{equation}
|V|=|J{\cal J}_{0}(K/\omega)|.\label{eq:PTCond1}
\end{equation}
When $|V|<|J{\cal J}_{0}(K/\omega)|$, the hopping process dominates
and all Floquet eigenstates of $\hat{H}_{1{\rm F}}$ are extended
with vanishing Lyapunov exponents (diverging localization lengths).
The quasienergy dispersion of $\hat{H}_{1{\rm F}}$ in this case takes
the form of $E(k)=2J{\cal J}_{0}\left(K/\omega\right)\cos(k)\in\mathbb{R}$,
where $k\in[0,2\pi)$. When $|V|>|J{\cal J}_{0}(K/\omega)|$, the
quasiperiodic potential dominates, and all Floquet eigenstates
of $\hat{H}_{1{\rm F}}$ are localized with the same Lyapunov exponent~(inverse localization length) $\lambda=-\ln|J{\cal J}_{0}(K/\omega)/V|>0$.
The Floquet spectrum of the system in this case has the form of $E(k)=2J{\cal J}_{0}\left(K/\omega\right)\cos(k-i\lambda)\in\mathbb{C}$,
where $k\in[0,2\pi)$. The extended and localized phases could be further
distinguished by a topological winding number of the Floquet spectrum.
With the eigenvalue equation~(\ref{eq:H1M}), we find
the matrix elements of the system's Floquet effective Hamiltonian $\hat{H}'_{1{\rm F}}$ 
in momentum space to be
\begin{alignat}{1}
[\hat{H}'_{1{\rm F}}]_{n,n} & =2J{\cal J}_{0}\left(\frac{K}{\omega}\right)\cos(2\pi\alpha n)\qquad n=1,...,L;\nonumber \\{}
[\hat{H}'_{1{\rm F}}]_{n+1,n} & =[\hat{H}'_{1{\rm F}}]_{1,L}=V,\qquad n=1,...,L-1.
\end{alignat}
Multiplying the corner matrix element $[\hat{H}'_{1{\rm F}}]_{1,L}$
by a phase factor $e^{-i\theta}$, we obtain the effective Hamiltonian
$\hat{H}'_{1{\rm F}}(\theta)$ under the twist boundary condition 
in momentum space. The spectral winding number of $\hat{H}'_{1{\rm F}}(\theta)$
can then be obtained following the construction of Ref.~\cite{ChenQC1}. Such
a winding number counts the number of times the Floquet spectrum of
$\hat{H}'_{1{\rm F}}(\theta)$ winds around a base quasienergy $E_{0}$
on the complex plane ${\rm Re}E$-${\rm Im}E$ when $\theta$
changes from zero to $2\pi$. When the quasienergy spectrum is real
(complex), the value of this winding number is found to be zero ($-1$).
This winding number could thus serve as a topological order parameter
to distinguish the extended phase (with real quasienergies) and localized
phase (with complex quasienergies) of the system. In Table~\ref{tab:ResM1},
we summarize the key results about the spectrum, localization transition
and topological invariant of the non-Hermitian Floquet quasicrystal
M1. Notably, the condition of localization transition in the
system now depends on the parameter $K/\omega$ of the driving field.
The states of the system can thus be tuned between extended and localized
phases dynamically by varying the ratio $K/\omega$ between the amplitude
and frequency of the driving force.

\begin{figure}[b]
	\begin{centering}
		\includegraphics[scale=0.45]{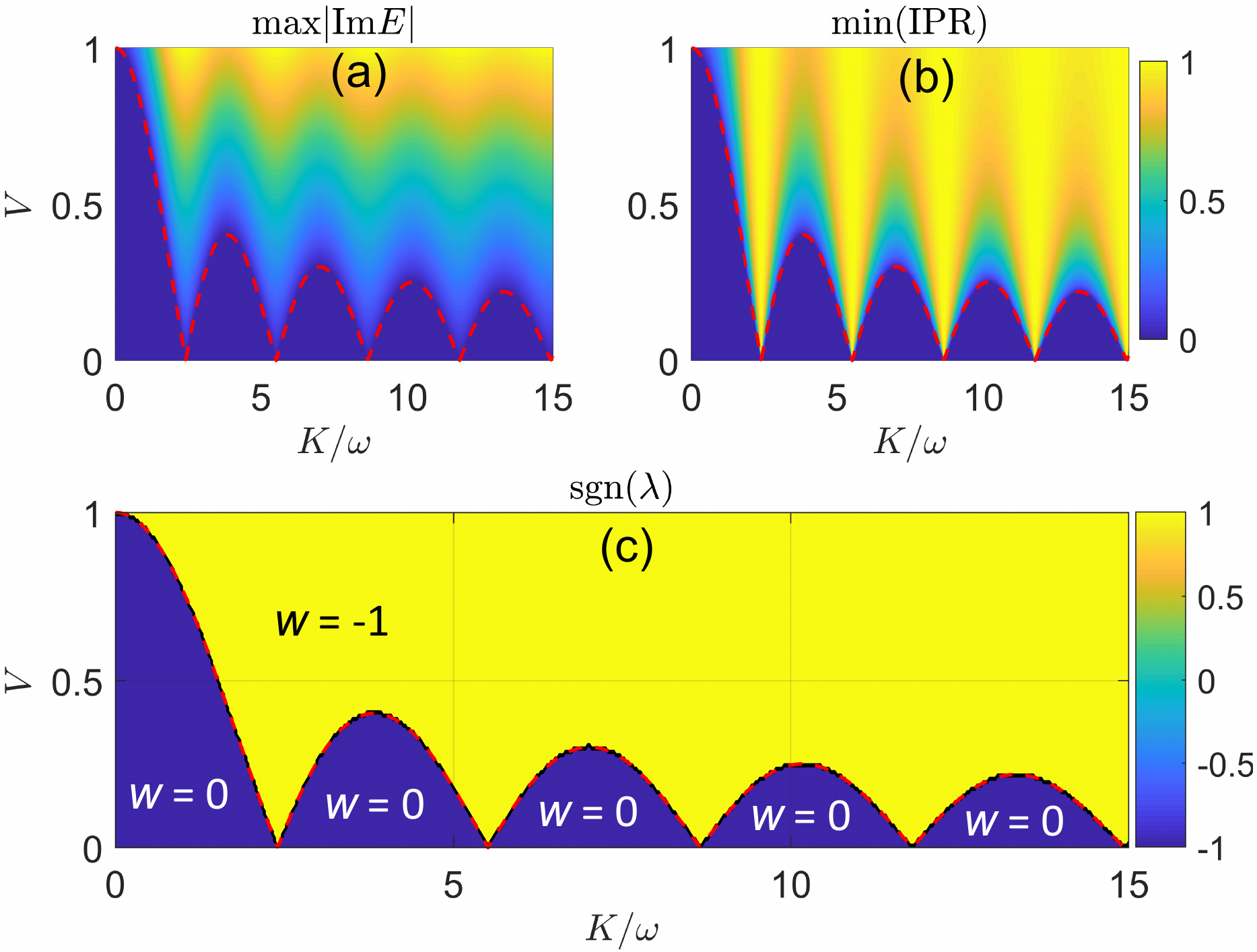}
		\par\end{centering}
	\caption{The maximal imaginary parts of Floquet quasienergies in panel (a),
		the minimal values of IPRs in panel (b), and the signs of Lyapunov
		exponents in panel (c) of the non-Hermitian Floquet quasicrystal M1.
		Other system parameters are set as $\alpha=\frac{\sqrt{5}-1}{2}$,
		$J=1$, and the length of lattice is taken as $L=610$. The red dashed
		line in each figure panel denotes the boundary between the extended
		phase (with real quasienergies for all Floquet eigenstates) and the
		localized phase (with complex quasienergies for all Floquet eigenstates),
		which satisfies the condition $|V|=|J{\cal J}_{0}(K/\omega)|$ (see
		Table~\ref{tab:ResM1}). The topological winding number $w$ of each
		phase is denoted explicitly in panel (c).\label{fig:M1}}
\end{figure}

To demonstrate our theoretical results, we compute the spectrum $E$,
inverse participation ratio (IPR), Lyapunov exponent $\lambda$ and
winding number $w$ of the M1 versus the amplitude of onsite potential
$V$ and the ratio between driving amplitude and frequency $K/\omega$.
The numerical results are presented in Fig.~\ref{fig:M1}. In the
calculation, we choose the undressed hopping amplitude $J=1$
and the quasicrystal parameter $\alpha=\frac{\sqrt{5}-1}{2}\simeq\frac{p}{q}$,
with $p,q$ being two adjacent elements of the Fibonacci sequence
($p<q$). The length of lattice is chosen as $L=610$, and the base quasienergy is set as $E_0=0$.  In Fig.~\ref{fig:M1}(a),
we show the maximal imaginary parts of quasienergy $\max|{\rm Im}E|$
versus $V$ and $K/\omega$. The red dashed line is given by the phase
boundary equation $|V|=|J{\cal J}_{0}(K/\omega)|$. It is clear that
$\max|{\rm Im}E|=0$ ($\max|{\rm Im}E|>0$) when $|V|<|J{\cal J}_{0}(K/\omega)|$
($|V|>|J{\cal J}_{0}(K/\omega)|$). The condition $|V|=|J{\cal J}_{0}(K/\omega)|$
thus determines the boundary of the ${\cal PT}$-transition between
real and complex quasienergy spectrum of the system. 

For any normalized
Floquet right eigenstate $|\psi\rangle=\sum_{n}\psi_{n}\hat{c}_{n}^{\dagger}|0\rangle$
of $\hat{H}_{1{\rm F}}$, with $\sum_{n=1}^{L}|\psi_{n}|^{2}=1$, we
define its IPR as
\begin{equation}
{\rm IPR}=\sum_{n=1}^{L}|\psi_{n}|^{4}.\label{eq:IPR}
\end{equation}
The minimum of IPRs over all eigenstates at a fix set of system parameters under the PBC
then yields the $\min({\rm IPR})$. In Fig.~\ref{fig:M1}(b), we present the $\min({\rm IPR})$ versus $V$ and
$K/\omega$, which is found to be $\simeq0$ when $|V|<|J{\cal J}_{0}(K/\omega)|$
(below the red dashed line) and $>0$ otherwise. The maximum
of IPRs of all eigenstates is found to have the same dependence on
$V$ and $K/\omega$. Therefore, the system is indeed in the extended
(localized) phase with all states being delocalized (localized) when
$|V|<|J{\cal J}_{0}(K/\omega)|$ ($|V|>|J{\cal J}_{0}(K/\omega)|$).
Finally, we present the signs of Lyapunov exponent $\lambda=-\ln|J{\cal J}_{0}(K/\omega)/V|$
and the topological winding numbers $w$ of M1 versus $V$ and $K/\omega$
in Fig.~\ref{fig:M1}(c). It is clear that the winding number $w=0$
in the extended phase with real quasieneriges, and $w=-1$ in the localized
phase with complex quasienergies. The two regions are separated by the phase
boundary $|V|=|J{\cal J}_{0}(K/\omega)|$ (red dashed line). Therefore,
it is verified that $w$ could be utilized as a topological order
parameter to discriminate the ${\cal PT}$-invariant extended phase
and ${\cal PT}$-breaking localized phase of the non-Hermitian Floquet
quasicrystal M1. Note that without the driving field, the phase boundary
between the extended and localized states reduces to a point at $V=J$.
The driving force instead allows the localized phase to disappear
and reappear alternately with the change of the driving parameter
$K/\omega$, which highlights the advantage of Floquet engineering
in the realization and control of phase transitions in non-Hermitian
quasicrystals.

\subsection{M2: Floquet spectrum, localization transition and topological invariant\label{subsec:M2}}
We next consider a driven non-Hermitian AAH quasicrystal with an imaginary
phase shift $i\gamma$, whose Hamiltonian takes the form of
\begin{alignat}{1}
\hat{H}_{2}(t)= & \sum_{n}J\left(\hat{c}_{n}^{\dagger}\hat{c}_{n+1}+{\rm h.c.}\right)\nonumber \\
+ & \sum_{n}\left[V\cos(2\pi\alpha n+i\gamma)-nK\cos(\omega t)\right]\hat{c}_{n}^{\dagger}\hat{c}_{n}.\label{eq:H2t}
\end{alignat}
According to the theory presented in Sec.~\ref{sec:Met}, the Floquet
effective Hamiltonian of this non-Hermitian quasicrystal M2 is
given by
\begin{alignat}{1}
\hat{H}_{2{\rm F}}= & \sum_{n}J{\cal J}_{0}\left(\frac{K}{\omega}\right)\left(\hat{c}_{n}^{\dagger}\hat{c}_{n+1}+{\rm h.c.}\right)\nonumber \\
+ & \sum_{n}V\cos(2\pi\alpha n+i\gamma)\hat{c}_{n}^{\dagger}\hat{c}_{n}.\label{eq:H2F}
\end{alignat}
It is clear that the hopping amplitude is modified by the driving
field and its magnitude could be controlled by the ratio between the
amplitude and frequency of the driving force $K/\omega$. Besides,
the $\hat{H}_{2{\rm F}}$ also holds the ${\cal PT}$-symmetry, since
its onsite potential $V_{n}=V\cos(2\pi\alpha n+i\gamma)=V_{-n}^{*}$.
The Floquet quasienergy spectrum of $\hat{H}_{2{\rm F}}$ could then
take real values in certain parameter regions. 

Plugging the single-particle state $|\psi\rangle=\sum_{n}\psi_{n}\hat{c}_{n}^{\dagger}|0\rangle$
into $\hat{H}_{2{\rm F}}|\psi\rangle=E|\psi\rangle$, we obtain
\begin{equation}
J{\cal J}_{0}\left(\frac{K}{\omega}\right)(\psi_{n+1}+\psi_{n-1})+V\cos(2\pi\alpha n+i\gamma)\psi_{n}=E\psi_{n},\label{eq:H2E}
\end{equation}
where $E\mod2\pi$ refers to the quasienergy. Taking the PBC for
a finite lattice of length $L$ and performing the discrete Fourier
transformation by Eq.~(\ref{eq:FFT}), we find the eigenvalue equation
in momentum space to be
\begin{equation}
\frac{V}{2}(e^{\gamma}\varphi_{n-1}+e^{-\gamma}\varphi_{n+1})+2J{\cal J}_{0}\left(\frac{K}{\omega}\right)\cos(2\pi\alpha n)\varphi_{n}=E\varphi_{n},\label{eq:H2M}
\end{equation}
which can also be viewed formally as a quasiperiodic lattice with
nonreciprocal hopping and ``photon-dressed'' onsite potential. 

\begin{table*}
	\begin{centering}
		\begin{tabular}{|c|c|c|}
			\hline 
			Phase & Extended & Localized\tabularnewline
			\hline 
			\hline 
			Condition & $|V|e^{|\gamma|}<|2J{\cal J}_{0}(K/\omega)|$ & $|V|e^{|\gamma|}>|2J{\cal J}_{0}(K/\omega)|$\tabularnewline
			\hline 
			IPR & $\simeq0$ for all states & $>0$ for all states\tabularnewline
			\hline 
			Lyapunov exponent & \multicolumn{2}{c|}{$\lambda=\ln\left|\frac{Ve^{|\gamma|}}{2J{\cal J}_{0}(K/\omega)}\right|\begin{cases}
				<0 & {\rm Extended}\\
				>0 & {\rm Localized}
				\end{cases}$}\tabularnewline
			\hline 
			Quasienergy & Real & Complex\tabularnewline
			\hline 
			Winding number & \multicolumn{2}{c|}{$w=\int_{0}^{2\pi}\frac{d\theta}{2\pi i}\partial_{\theta}\ln\det[\hat{H}_{2{\rm F}}(\theta)-E_{0}]=\begin{cases}
				0 & {\rm Extended}\\
				-1 & {\rm Localized}
				\end{cases}$}\tabularnewline
			\hline 
		\end{tabular}
		\par\end{centering}
	\caption{Summary of the results for the non-Hermitian Floquet quasicrystal M2.
		$J$ is the nearest-neighbor hopping amplitude, $V$ is the amplitude
		of onsite potential, $\gamma$ is the imaginary part
		of superlattice phase shift, $K$ is the driving amplitude, $\omega$
		is the driving frequency, and ${\cal J}_{0}(K/\omega)$ denotes the
		Bessel function of first kind. $\lambda$ denotes the Lyapunov exponent~\cite{Note1}. $H_{2{\rm F}}(\theta)$ is obtained
		from $H_{2{\rm F}}$ after setting $V_{n}=V\cos(2\pi\alpha n+i\gamma+\theta/L)$,
		with $L$ being the length of lattice. $E_{0}$ is the base quasienergy.\label{tab:ResM2}}
\end{table*}

Following the method developed in Ref.~\cite{LonghiQC1}, it is straightforward
to show that the non-Hermitian Floquet quasicrystal M2 could undergo
a ${\cal PT}$-transition and a localization-delocalization transition
when its parameters satisfy the condition
\begin{equation}
|V|e^{|\gamma|}=|2J{\cal J}_{0}(K/\omega)|,\label{eq:PTCond2}
\end{equation}
which is clearly dependent on the parameters of the driving field
$K$ and $\omega$. If $|V|e^{|\gamma|}<|2J{\cal J}_{0}(K/\omega)|$,
the hopping energy overwhelms the quasiperiodic onsite disorder, and
the system sits in an extended phase. All Floquet eigenstates
in this phase have real quasieneriges and delocalized profiles
(with vanishing IPRs) throughout the lattice. If $|V|e^{|\gamma|}>|2J{\cal J}_{0}(K/\omega)|$,
the non-Hermitian onsite potential governs the behavior of the system
and drives it into a localized phase. The Floquet states of the system
in this phase are all localized (with finite IPRs) and have complex
quasienergies. These localized eigenstates are found to share the
common Lyapunov exponent $\lambda=-\ln|2J{\cal J}_{0}(K/\omega)/(Ve^{|\gamma|})|$,
whose value is also controlled by the ratio between the amplitude
and frequency of the driving force. To construct a topological winding
number for M2, we add a real phase shift to the quasiperiodic
onsite potential~\cite{LonghiQC1} by setting $\cos(2\pi\alpha n+i\gamma)\rightarrow\cos(2\pi\alpha n+i\gamma+\theta/L)$
in Eq.~(\ref{eq:H2F}), where $L$ is the length of lattice. We denote
the phase-shifted effective Hamiltonian by $\hat{H}_{2{\rm F}}(\theta)$,
and the number of times that its quasienergy spectrum winds around
a base point $E_{0}$ on the complex plane defines a topological
invariant of the system, which is equal to $-1$ when the spectrum
is complex, and zero otherwise. Therefore, the non-Hermitian
Floquet quasicrystal M1 possesses an extended phase with real spectrum
and winding number $w=0$, and a localized phase with complex spectrum
and winding number $w=-1$. By tuning the ratio $K/\omega$ between
the amplitude and frequency of the driving force, the system could
also roam alternately between the ${\cal PT}$-invariant extended
phase and the ${\cal PT}$-breaking localized phase. The
topological and transport properties of the non-Hermitian quasicrystal
M2 could thus be controlled by the periodic driving field. We summarize
the key results about the spectrum, phases and topological features
of M2 in Table~\ref{tab:ResM2}.

\begin{figure}
	\begin{centering}
		\includegraphics[scale=0.45]{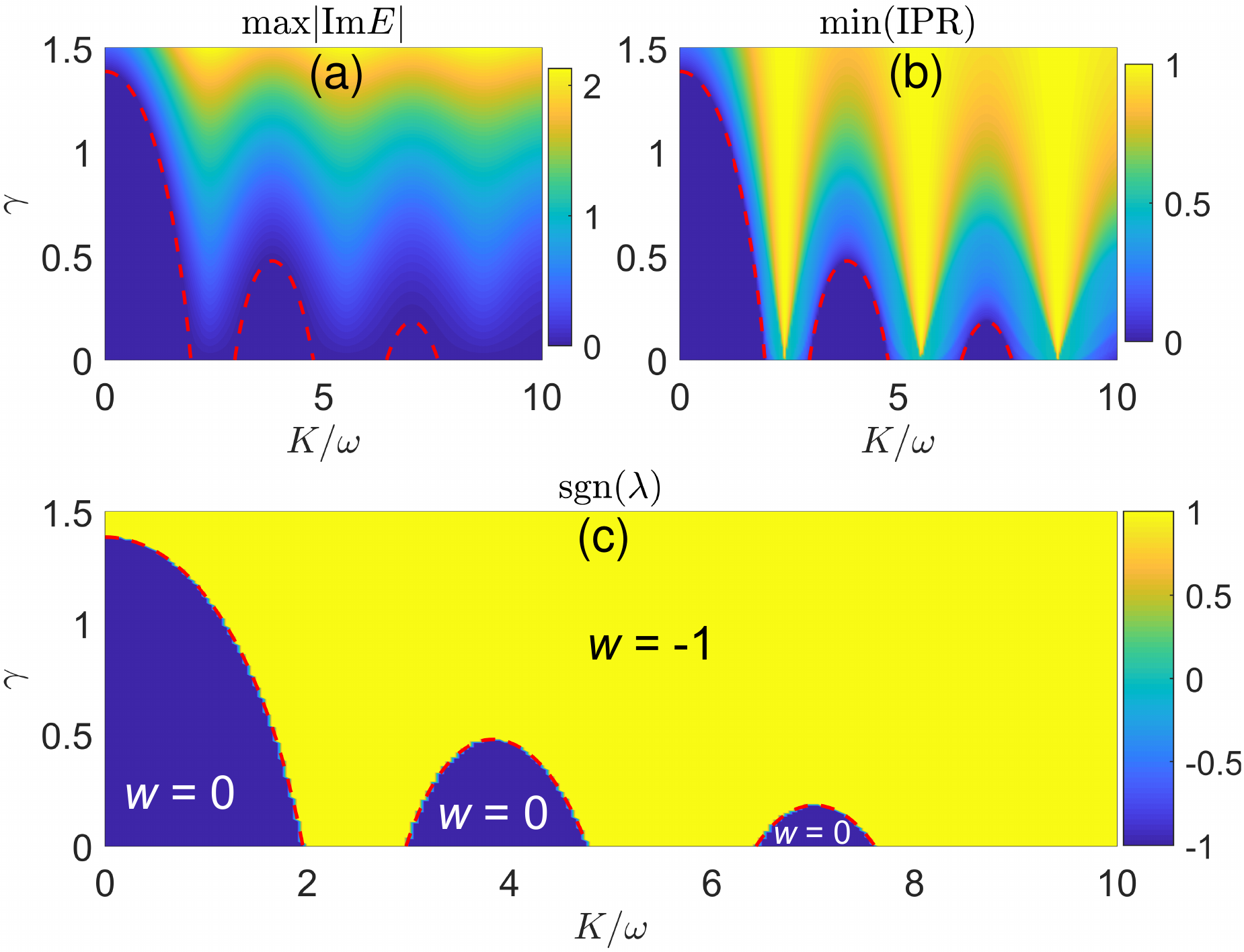}
		\par\end{centering}
	\caption{The maximal imaginary parts of Floquet quasienergies in panel (a),
		the minimal values of IPRs in panel (b), and the signs of Lyapunov
		exponents in panel (c) of the non-Hermitian Floquet quasicrystal M2.
		Other system parameters are set as $\alpha=\frac{\sqrt{5}-1}{2}$,
		$J=2$, $V=1$, and the length of lattice is taken as $L=610$. The
		red dashed line in each figure panel denotes the boundary between
		the extended phase (with real quasienergies for all Floquet eigenstates)
		and the localized phase (with complex quasienergies for all states), which satisfies the condition $|V|e^{|\gamma|}=|J{\cal J}_{0}(K/\omega)|$
		(see Table \ref{tab:ResM2}). The topological winding number $w$
		of each phase is denoted explicitly in panel (c).\label{fig:M2}}
\end{figure}

To verify our theoretical predictions, we compute the quasienergy
spectrum, IPR, Lyapunov exponent and winding number of the non-Hermitian
Floquet quasicrystal M2 numerically, with results reported in Fig.~\ref{fig:M2}. In the calculation, we choose the system parameters
$(J,V)=(2,1)$ and the quasiperiodic parameter $\alpha=\frac{\sqrt{5}-1}{2}\simeq\frac{p}{q}$,
where $p,q$ are two neighboring elements of the Fibonacci sequence
($p<q$). The length of lattice is set as $L=610$, and the base quasienergy is chosen to be $E_0=0$. In Fig.~\ref{fig:M2}(a),
we show the maximum of the imaginary parts of Floquet spectrum versus
the parameter of driving force $K/\omega$ and imaginary part of phase
shift $\gamma$. The red-dashed line highlights the boundary of ${\cal PT}$-transition,
which is determined by Eq.~(\ref{eq:PTCond2}). The numerical results
confirm that the Floquet spectrum of the system is indeed real when
$|V|e^{|\gamma|}<|2J{\cal J}_{0}(K/\omega)|$ (below the red-dashed
line), and complex in other circumstances. In Fig.~\ref{fig:M2}(b),
we show the minimum of IPRs, which behaves similarly as the maximum
of IPRs versus $K/\omega$ and $\gamma$ at each given set of system
parameters. The results indicate that the IPRs of all states approach
zero when $|V|e^{|\gamma|}<|2J{\cal J}_{0}(K/\omega)|$ (below
the red-dashed line), and become finite if $|V|e^{|\gamma|}>|2J{\cal J}_{0}(K/\omega)|$.
Therefore, the condition in Eq.~(\ref{eq:PTCond2}) defines the boundary
between the ${\cal PT}$-invariant extended phase and the ${\cal PT}$-breaking
localized phase of the non-Hermitian Floquet quasicrystal M2. Finally,
we show the signs of Lyapunov exponent $\lambda$ and the
topological winding numbers $w$ versus $K/\omega$ and $\gamma$ in
Fig.~\ref{fig:M2}(c), in which we observe $\lambda<0$ ($\lambda>0$)
for all states in the extended (localized) phase with $w=0$ ($w=-1$),
as shown by the blue (yellow) regions below (above) the phase boundary
(i.e., the red-dashed line satisfying Eq.~(\ref{eq:PTCond2})). Therefore,
the winding number $w$ works as a topological invariant to distinguish
the ${\cal PT}$-breaking localized phase and ${\cal PT}$-invariant
extended phase of the system. Similar to the case of M1, the driving
force now allows one to change the system over a series of transitions
between localized and delocalized phases. This further demonstrates
the generality of our Floquet engineering approach to the design and
control of non-Hermitian quasicrystalline phases in different model
systems. 

\subsection{M3: Floquet spectrum, localization transition and topological invariant\label{subsec:M3}}

\begin{table*}
	\begin{centering}
		\begin{tabular}{|c|c|c|}
			\hline 
			Phase & Extended & Localized\tabularnewline
			\hline 
			\hline 
			Condition & $|V|e^{-|\gamma|}<|2J{\cal J}_{0}(K/\omega)|$ & $|V|e^{-|\gamma|}>|2J{\cal J}_{0}(K/\omega)|$\tabularnewline
			\hline 
			IPR & $\simeq0$ for all states & $>0$ for all states\tabularnewline
			\hline 
			Lyapunov exponent & \multicolumn{2}{c|}{$\lambda=\ln\left|\frac{Ve^{-|\gamma|}}{2J{\cal J}_{0}(K/\omega)}\right|\begin{cases}
				<0 & {\rm Extended}\\
				>0 & {\rm Localized}
				\end{cases}$}\tabularnewline
			\hline 
			Quasienergy & Complex & Real\tabularnewline
			\hline 
			Winding number & \multicolumn{2}{c|}{$w=\int_{0}^{2\pi}\frac{d\theta}{2\pi i}\partial_{\theta}\ln\det[H_{3{\rm F}}(\theta)-E_{0}]=\begin{cases}
				-1 & {\rm Extended}\\
				0 & {\rm Localized}
				\end{cases}$}\tabularnewline
			\hline 
		\end{tabular}
		\par\end{centering}
	\caption{Summary of results for the non-Hermitian Floquet quasicrystal M3.
		$J$ is the symmetric part of nearest-neighbor hopping amplitude,
		$\gamma$ controls the asymmetry between left and right hopping amplitudes,
		$V$ is the amplitude of onsite quasiperiodic potential, $K$ is the
		driving amplitude, $\omega$ is the driving frequency, and ${\cal J}_{0}(K/\omega)$
		denotes the Bessel function of first kind. $\lambda$ denotes the Lyapunov exponent~\cite{Note1}. $H_{3{\rm F}}(\theta)$
		is obtained from $H_{3{\rm F}}$ after taking the periodic boundary
		condition and setting its top-right (bottom-left) corner matrix element
		to be $J_{L}{\cal J}_0(K/\omega)e^{-i\theta}$ ($J_{R}{\cal J}_0(K/\omega)e^{i\theta}$). $E_{0}$ is the
		base quasienergy.\label{tab:ResM3}}
\end{table*}

In the thrid part of this section, we consider a nonreciprocal variant
of the AAH model~\cite{LonghiQC4,ChenQC1}, which is subject to a periodically modulated driving
force. The time-dependent Hamiltonian of the model takes the form
\begin{alignat}{1}
\hat{H}_{3}(t)= & \sum_{n}J\left(e^{-\gamma}\hat{c}_{n}^{\dagger}\hat{c}_{n+1}+e^{\gamma}\hat{c}_{n+1}^{\dagger}\hat{c}_{n}\right)\nonumber \\
+ & \sum_{n}\left[V\cos(2\pi\alpha n)-nK\cos(\omega t)\right]\hat{c}_{n}^{\dagger}\hat{c}_{n},\label{eq:H3t}
\end{alignat}
which is non-Hermitian if $\gamma\neq0$. Following the steps of Sec.~\ref{sec:Met}, we obtain the Floquet effective Hamiltonian of this
non-Hermitian quasicrystal M3 under the high-frequency approximation
as
\begin{alignat}{1}
\hat{H}_{3{\rm F}}= & \sum_{n}{\cal J}_{0}\left(\frac{K}{\omega}\right)J\left(e^{-\gamma}\hat{c}_{n}^{\dagger}\hat{c}_{n+1}+e^{\gamma}\hat{c}_{n+1}^{\dagger}\hat{c}_{n}\right)\nonumber \\
+ & \sum_{n}V\cos(2\pi\alpha n)\hat{c}_{n}^{\dagger}\hat{c}_{n}.\label{eq:H3F}
\end{alignat}
The hopping amplitude is now controlled by the amplitude $K$ and
frequency $\omega$ of the driving field. Applying $\hat{H}_{3{\rm F}}$
to the eigenstate $|\psi\rangle=\sum_{n}\psi_{n}\hat{c}_{n}^{\dagger}|0\rangle$,
we further obtain the eigenvalue equation in lattice representation
\begin{equation}
{\cal J}_{0}\left(\frac{K}{\omega}\right)J(e^{\gamma}\psi_{n-1}+e^{-\gamma}\psi_{n+1})+V\cos(2\pi\alpha n)\psi_{n}=E\psi_{n}.\label{eq:H3E}
\end{equation}
That this equation takes the same form as Eq.~(\ref{eq:H2M})
under the exchange of system parameters $V\leftrightarrow2J{\cal J}_{0}(K/\omega)$,
which implies that the non-Hermitian Floquet quasicrystals M2 and
M3 are dual to each other concerning their spectrum and states. Furthermore, taking the Fourier transformation
of Eq.~(\ref{eq:H3E}) with the help of Eq.~(\ref{eq:FFT}), we obtain
\begin{equation}
\frac{V}{2}(\varphi_{n+1}+\varphi_{n-1})+2J{\cal J}_{0}\left(\frac{K}{\omega}\right)\cos(2\pi\alpha n-i\gamma)\varphi_{n}=E\varphi_{n},\label{eq:H3M}
\end{equation}
which again shares the same form with Eq.~(\ref{eq:H2E}).
Putting together, we find that under the PBC, there is indeed a duality
relation between the M2 and M3, in the sense that the localization properties of the M2 in momentum space is consistent
with the localization properties of the M3 in position
space under the PBC. With the help of this duality relation, we can immediately
write down the phase boundary condition that determines the ${\cal PT}$-transition
and localization-delocalization transition in the M3 under the PBC, i.e., 
\begin{equation}
|V|=|2J{\cal J}_{0}(K/\omega)|e^{|\gamma|}.\label{eq:PTCond3}
\end{equation}
In parallel with the discussions of Subsec.~\ref{subsec:M2}, if $|V|<|2J{\cal J}_{0}(K/\omega)|e^{|\gamma|}$,
the M3 is found to be in a ${\cal PT}$-breaking extended phase (dual
to the localized phase of M2) with complex quasienergies and vanishing
IPRs for all Floquet eigenstates. When $|V|>|2J{\cal J}_{0}(K/\omega)|e^{|\gamma|}$,
the M3 is in a ${\cal PT}$-invariant localized phase (dual to the
extended phase of M2), where all Floquet eigenstates have real quasienergies,
finite IPRs, and the same positive Lyapunov exponent $\lambda=-\ln|2J{\cal J}_{0}(K/\omega)e^{|\gamma|}/V|$
in the thermodynamic limit $L\rightarrow\infty$. Note that due to
the dependence of phase boundary condition Eq.~(\ref{eq:PTCond3})
and Lyapunov exponent on the ratio between the amplitude and frequency
of the driving force $K/\omega$, transitions between phases with
different spectrum and transport features in the M3 can also be induced
by the driving field. 

To define a spectral winding number, we take
the twist boundary condition~\cite{ChenQC1} by adding phase factors to the corner
matrix elements of $\hat{H}_{3{\rm F}}$ in the lattice representation,
i.e., 
\begin{equation}
[\hat{H}_{3{\rm F}}]_{1,L}=Je^{(\gamma-i\theta)},\qquad[\hat{H}_{3{\rm F}}]_{L,1}=Je^{(i\theta-\gamma)},
\end{equation}
where $L$ is the length of lattice and $\theta\in[0,2\pi)$. The
spectrum of the phase-dependent Hamiltonian $\hat{H}_{3{\rm F}}(\theta)$
could now possess a winding number around certain base quasienergy
$E_0$ on the complex plane when $\theta$ changes over a cycle from zero
to $2\pi$. Such a winding number could be nonzero only if the
spectrum of $\hat{H}_{3{\rm F}}(\theta)$ is complex. Therefore, we
expect a quantized winding number to appear for the ${\cal PT}$-breaking
extended phase, and a vanishing one for the ${\cal PT}$-invariant
localized phase of the system. Such a winding number can then be used
as a topological invariant to distinguish the two non-Hermitian Floquet
quasicrystalline phases of M3 with distinct spectrum and transport
nature. We summarize the main results about the M3 in Table~\ref{tab:ResM3}
for the ease of reference. 

\begin{figure}
	\begin{centering}
		\includegraphics[scale=0.45]{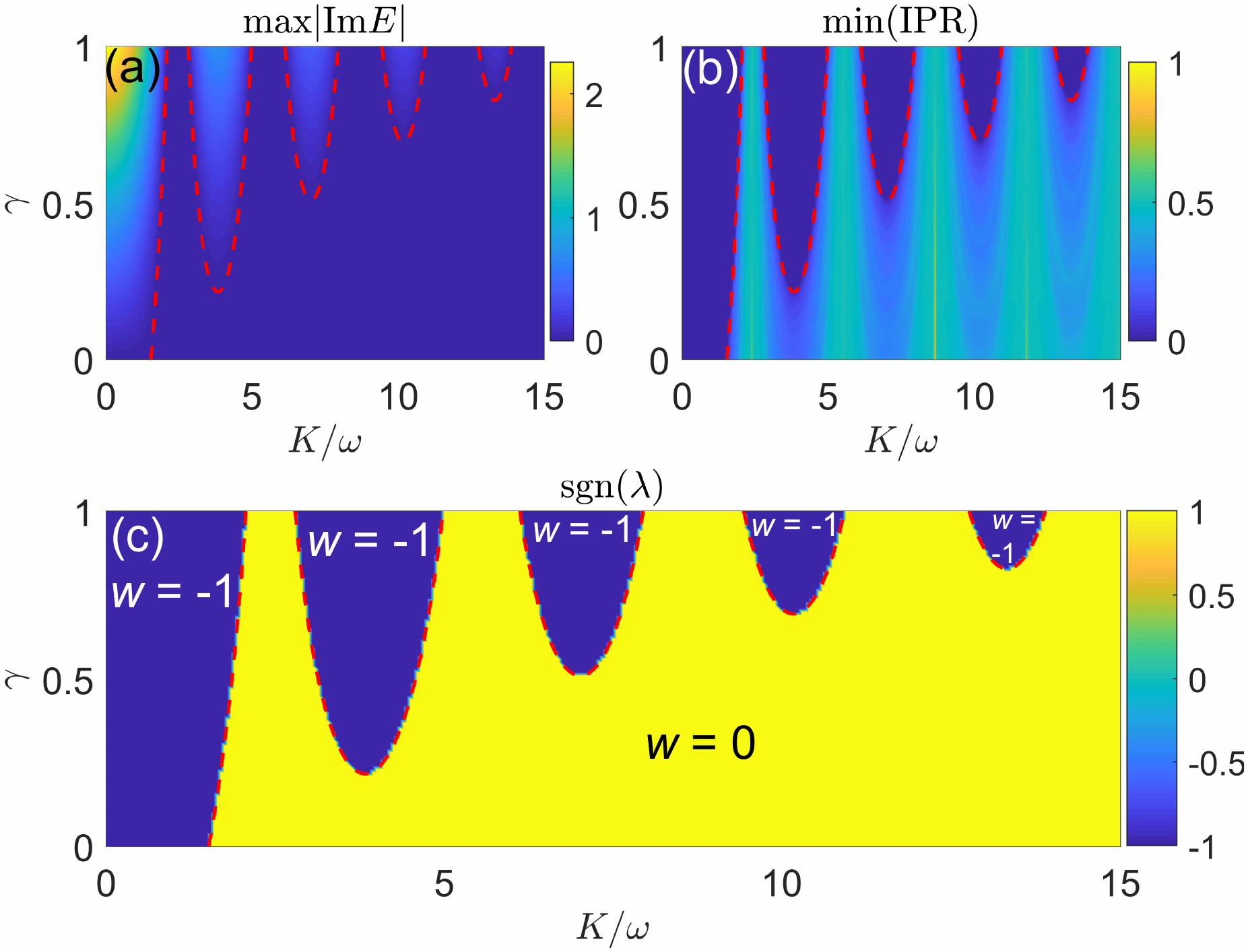}
		\par\end{centering}
	\caption{The maximal imaginary parts of quasienergies in panel (a),
		the minimal values of IPRs in panel (b), and the signs of Lyapunov
		exponents in panel (c) of the non-Hermitian Floquet quasicrystal M3.
		Other system parameters are set as $\alpha=\frac{\sqrt{5}-1}{2}$,
		$J=1$, $V=1$, and the length of lattice is taken as $L=610$. The
		red dashed line in each figure panel denotes the boundary between
		the extended phase (with complex quasienergies for all states)
		and the localized phase (with real quasienergies),
		which satisfies the condition $|V|e^{-|\gamma|}=|J{\cal J}_{0}(K/\omega)|$
		(see Table \ref{tab:ResM3}). The topological winding number $w$
		of each phase is denoted explicitly in panel (c).\label{fig:M3}}
\end{figure}

To confirm our theoretical predictions, we present the Floquet spectrum,
IPR, Lyapunov exponent and winding number of the M3 versus the imaginary
part of phase shift $\gamma$ and the ratio between the amplitude
and frequency of driving field $K/\omega$ in Fig.~\ref{fig:M3}.
Other system parameters are set as $J=V=1$ in the calculation. The
quasicrystal parameter $\alpha=\frac{\sqrt{5}-1}{2}\simeq\frac{p}{q}$,
with $p,q$ being adjacent elements of the Fibonacci sequence ($p<q$).
The length of lattice is chosen to be $L=610$, and the base quasienergy is set as $E_0=0$. In Figs.~\ref{fig:M3}(a)\textendash (c),
the red dashed lines refer to the phase boundary satisfying the Eq.~(\ref{eq:PTCond3}). We observe that the maximum of imaginary parts
of quasienergy $\max|{\rm Im}E|$ vanishes below the red dashed line
and taking finite values above it, as shown in Fig.~\ref{fig:M3}(a).
The system thus undergoes a ${\cal PT}$-transition when $|V|e^{-|\gamma|}$ increases
from below $|2J{\cal J}_{0}(K/\omega)|$ to above it,
through which the Floquet spectrum changes from real to complex. In
Fig.~\ref{fig:M3}(b), we show the minimum of IPRs 
over all Floquet eigenstates at each given set of system parameters.
The maximum of IPRs versus $K/\omega$ and $\gamma$ are found to
have the same pattern. Therefore, all states of the system are extended
in the ${\cal PT}$-breaking regime (with complex spectrum), and localized
in the ${\cal PT}$-invariant regime (with real spectrum). In Fig.~\ref{fig:M3}(c), we show the signs of Lyapunov exponent ${\rm sgn}(\lambda)$
and winding numbers $w$ of different phases. We obtain identical
and positive (negative) $\lambda$ for all states below (above) the
phase boundary line, implying a phase with localized (delocalized)
states at all quasienergies. Moreover, we find the winding number
$w=0$ and $w=-1$ in the localized and delocalized phases, respectively,
and a quantized change of $w$ across the phase boundary (red dashed
line). The driven non-Hermitian quasicrystal M3 thus possesses a ${\cal PT}$-breaking
extended phase with topological winding number $w=-1$, in which all
Floquet eigenstates are delocalized with complex quasienergies, and
a ${\cal PT}$-invariant localized phase with topological winding
number $w=0$, in which all Floquet eigenstates are localized with
real quasienergies. As the phase boundary is controlled by the amplitude
and frequency of the driving force, the system can also be tuned dynamically
to go through transitions between different non-Hermitian quasicrystalline
phases following our Floquet engineering scheme. 

\subsection{M4: Floquet spectrum, localization transition, mobility edge and topological invariant\label{subsec:M4}}

\begin{table*}
	\begin{centering}
		\begin{tabular}{|c|c|c|c|}
			\hline 
			Phase & Localized & Mobility Edge & Extended\tabularnewline
			\hline 
			\hline 
			Condition & $\gamma<\gamma_{1}$ & $\gamma_{1}<\gamma<\gamma_{2}$ & $\gamma_{2}<\gamma$\tabularnewline
			\hline 
			IPR & $>0$ for all states & $>0$ \& $\simeq0$ coexist & $\simeq0$ for all states\tabularnewline
			\hline 
			Lyapunov exponent & \multicolumn{3}{c|}{$\lambda(E)={\rm arccosh}\left[\frac{\sqrt{[2J{\cal J}_{0}(K/\omega)+E_{{\rm r}}]^{2}+(V-E_{{\rm i}})^{2}}+\sqrt{[2J{\cal J}_{0}(K/\omega)-E_{{\rm r}}]^{2}+(V-E_{{\rm i}})^{2}}}{4|J{\cal J}_{0}(K/\omega)|}\right]\begin{cases}
				=0 & {\rm Extended}\\
				>0 & {\rm Localized}
				\end{cases}$}\tabularnewline
			\hline 
			Quasienergy & \multicolumn{3}{c|}{$E=\begin{cases}
				\pm\sqrt{[2J{\cal J}_{0}(K/\omega)\cos(\beta+i\gamma)]^{2}+V^{2}\cot^{2}(\beta+i\gamma)} & {\rm Localized\,\,States}\\
				2J{\cal J}_{0}(K/\omega)\cos\beta+iV & {\rm Extended\,\,States}
				\end{cases}$}\tabularnewline
			\hline 
			Winding number & \multicolumn{3}{c|}{$w_{1,2}=\int_{0}^{\pi}\frac{d\theta}{2\pi i}\partial_{\theta}\ln\det[H_{4{\rm F}}(\theta)-E_{1,2}]=\begin{cases}
				(1,1) & {\rm Localized}\\
				(0,1) & {\rm Mobility\,\,Edge}\\
				(0,0) & {\rm Extended}
				\end{cases}$}\tabularnewline
			\hline 
		\end{tabular}
		\par\end{centering}
	\caption{Summary of results for the non-Hermitian Floquet quasicrystal M4.
		$J$ is the nearest-neighbor hopping amplitude, $V$ is the amplitude
		of onsite potential, $\gamma$ is the imaginary part
		of superlattice phase shift, $K$ is the driving amplitude, $\omega$
		is the driving frequency, and ${\cal J}_{0}(K/\omega)$ denotes the
		Bessel function of first kind. The phase variable $\beta\in[-\pi,\pi)$.
		$H_{4{\rm F}}(\theta)$ is obtained from $H_{4{\rm F}}$ after setting
		$V_{n}=V\tan(\pi\alpha n+i\gamma+\theta/L)$, with $L$ being the
		length of lattice. $E_{1}=iV$ and $E_{2}=2J{\cal J}_{0}(K/\omega)+iV$
		are two base quasienergies with respect to which the two winding numbers
		$(w_{1},w_{2})$ are defined.\label{tab:ResM4}}
\end{table*}

To further reveal the usefulness of the Floquet engineering
scheme, we now apply it to a non-Hermitian quasicrystal that could possess energy-dependent
mobility edges. In the lattice representation, the Hamiltonian of
the system describes a non-Hermitian extension of the Maryland model~\cite{MM1,MM2,MM3}
subjecting to an oscillating force, which takes the form
\begin{alignat}{1}
\hat{H}_{4}(t)= & \sum_{n}J\left(\hat{c}_{n}^{\dagger}\hat{c}_{n+1}+{\rm h.c.}\right)\nonumber \\
+ & \sum_{n}\left[V\tan(\pi\alpha n+i\gamma)-nK\cos(\omega t)\right]\hat{c}_{n}^{\dagger}\hat{c}_{n}.\label{eq:H4t}
\end{alignat}
Following the theory presented in Sec.~\ref{sec:Met}, the Floquet
effective Hamiltonian of this non-Hermitian quasicrystal M4 reads
\begin{alignat}{1}
\hat{H}_{4{\rm F}}= & \sum_{n}J{\cal J}_{0}\left(\frac{K}{\omega}\right)\left(\hat{c}_{n}^{\dagger}\hat{c}_{n+1}+{\rm h.c.}\right)\nonumber \\
+ & \sum_{n}V\tan(\pi\alpha n+i\gamma)\hat{c}_{n}^{\dagger}\hat{c}_{n},\label{eq:H4F}
\end{alignat}
in which the fast driving field generates a modulation to the hopping
amplitude through the factor ${\cal J}_{0}(K/\omega)$. Expressing
the Floquet eigenstate as $|\psi\rangle=\sum_{n}\psi_{n}\hat{c}_{n}^{\dagger}|0\rangle$,
we obtain the eigenvalue equation
\begin{equation}
J{\cal J}_{0}\left(\frac{K}{\omega}\right)(\psi_{n+1}+\psi_{n-1})+V\tan(\pi\alpha n+i\gamma)\psi_{n}=E\psi_{n},\label{eq:H4E}
\end{equation}
where $E\mod2\pi$ refers to the quasienergy of $|\psi\rangle$. 

Following the mapping between the Maryland model and an equivalent Floquet problem~\cite{MM1,MM2,MM3}, it can
be shown that this non-Hermitian quasicrystal M4 possesses
two phase transitions at $\gamma=\gamma_{1}$ and $\gamma=\gamma_{2}$,
with
\begin{alignat}{1}
\gamma_{1} & =\frac{1}{2}{\rm arcsinh}\left(V/|J{\cal J}_{0}(K/\omega)|\right),\label{eq:GAM1}\\
\gamma_{2} & =\frac{1}{2}{\rm arccosh}\sqrt{\Delta+\sqrt{\Delta^{2}-1}},\label{eq:GAM2}
\end{alignat}
where $\Delta\equiv(2+[V/J{\cal J}_{0}(K/\omega)]^{2})/2$. When the
imaginary part of phase shift $\gamma<\gamma_{1}$, all eigenstates
of the system are localized with quasienergies
\begin{equation}
E=\pm\sqrt{[2J{\cal J}_{0}(K/\omega)\cos(\beta+i\gamma)]^{2}+V^{2}\cot^{2}(\beta+i\gamma)},\label{eq:M4E1}
\end{equation}
where $\beta\in[0,\pi]$ and ${\rm Im}E>0$. These quasienergies form
a closed loop around $E_{1}=iV$ on the complex plane, whose size
and shape can be controlled by the ratio between the amplitude and
frequency of the driving force. 

At $\gamma=\gamma_{1}$, the system undergoes a transition from a
localized phase to a phase with quasienergy-dependent mobility edges,
and stays in this mobility edge phase when $\gamma\in(\gamma_{1},\gamma_{2})$.
The spectrum of the system in this phase can be separated into
two parts. Within the range of $(-2|\cos\beta_{0}|,2|\cos\beta_{0}|)$
on the complex energy plane, where
\begin{equation}
\beta_{0}=\arccos\left(\cos(2\gamma)\sqrt{1-\frac{V^{2}}{[J{\cal J}_{0}(K/\omega)\sin(2\gamma)]^{2}}}\right),\label{eq:PHI0}
\end{equation}
the imaginary part of the spectrum is pinned at $V$, and $E$ takes
the form of
\begin{equation}
E=2J{\cal J}_{0}(K/\omega)\cos\beta+iV,\quad\beta\in[-|\beta_{0}|,|\beta_{0}|].\label{eq:M4E2}
\end{equation}
It is clear that the two ends of the spectrum along the real
axis are controlled by the driving field, which could merge into one
when $K/\omega$ is tuned to the zeros of the Bessel function ${\cal J}_{0}$.
Since only the states whose quasienergies satisfy Eq.~(\ref{eq:M4E2})
have delocalized profiles in the mobility edge phase, the driving
field provides us with a convenient knob to engineer and control the
mobility edges of the non-Hermitian Maryland model. Beyond the range
of $(-2|\cos\beta_{0}|,2|\cos\beta_{0}|)$, the spectrum of the system
is still described by Eq.~(\ref{eq:M4E1}), which now arranges into
two loops around $E_{2}^{\pm}=\pm2J{\cal J}_{0}(K/\omega)+iV$. 

At $\gamma=\gamma_{2}$, the system experiences a second transition
from the mobility edge phase to an extended phase, in which all eigenstates
are delocalized and their quasienergies are described by Eq.~(\ref{eq:M4E2})
for $\gamma>\gamma_{2}$, with $\beta\in[-\pi,\pi)$. The spectrum
now collapses into a line ending at $E_{2}^{\pm}$,
and the loops described by Eq.~(\ref{eq:M4E1}) all vanish. The ending
points of the spectrum of delocalized states could again be controlled
by the driving field.

Following the Avila\textquoteright s global theory~\cite{MM3,Avila1,Avila2}, the system
is found to have a quasienergy-dependent Lyapunov exponent $\lambda(E=E_{{\rm r}}+iE_{{\rm i}})$
in the thermodynamic limit, where $E_{{\rm r}}$ and $E_{{\rm i}}$
are the real and imaginary parts of $E$ {[}see Table
\ref{tab:ResM4} for the expression of $\lambda(E)${]}. In the extended
and localized phases, one has $\lambda(E)=0$ and $\lambda(E)>0$
for all states, respectively. In the mobility edge phase, we have
$\lambda(E)=0$ for eigenstates with $E_{{\rm r}}\in(-2|\cos\beta_{0}|,2|\cos\beta_{0}|)$,
and $\lambda(E)>0$ otherwise. As expected, the Lyapunov exponents,
and therefore the localization nature of the Floquet states can
be controlled by the ratio between the amplitude and frequency of
the driving field $K/\omega$.

\begin{figure}
	\begin{centering}
		\includegraphics[scale=0.46]{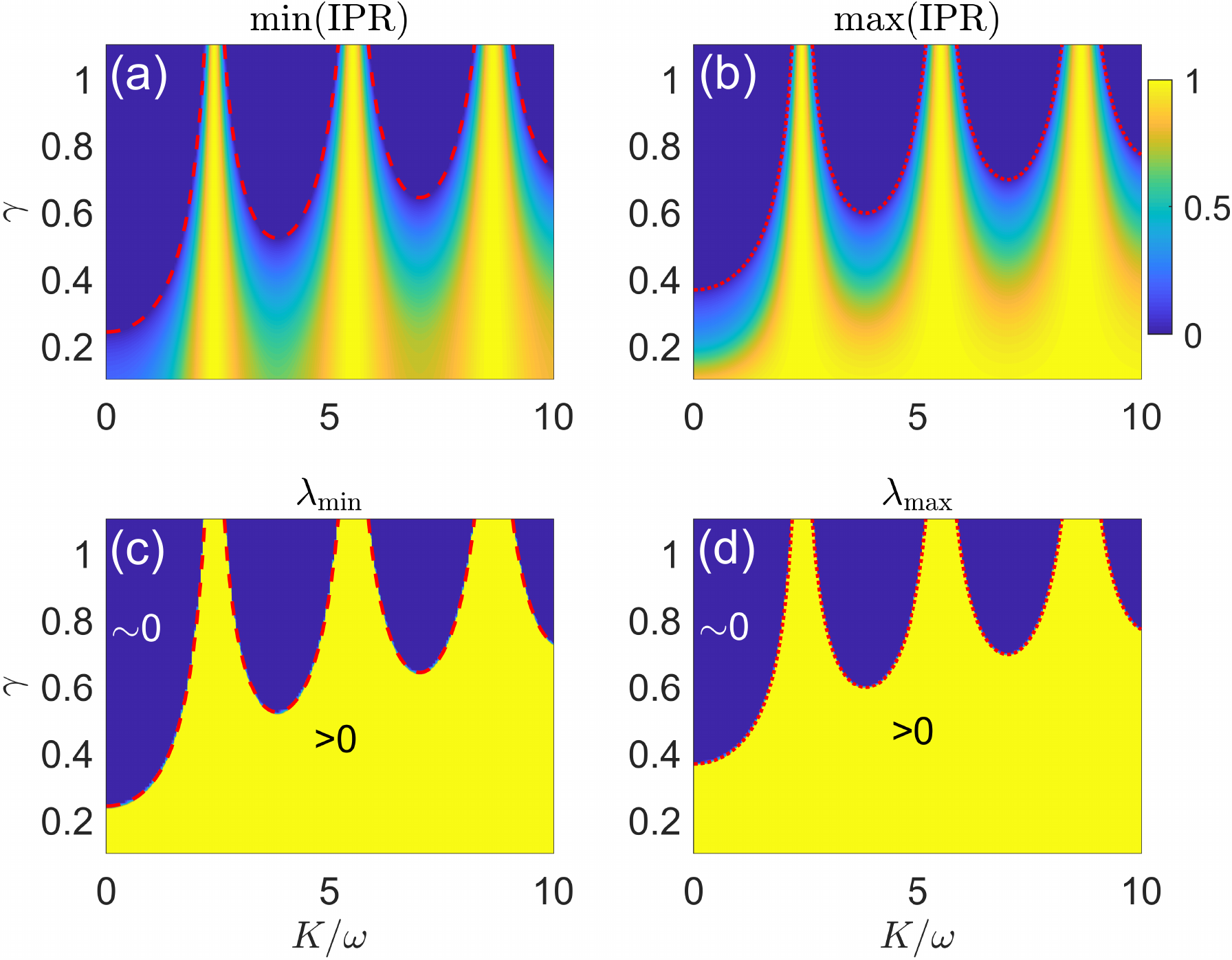}
		\par\end{centering}
	\caption{The the minimal (maximal) values of IPRs in panel (a)~[(b)], and the minimum (maximum) of Lyapunov
		exponents in panel (c)~[(d)] of the non-Hermitian Floquet quasicrystal M4.
		Other system parameters are set as $\alpha=\frac{\sqrt{5}-1}{2}$,
		$J=2$, $V=1$, and the length of lattice is taken as $L=610$. The
		red dashed and dotted lines in each figure panel denote the boundaries between
		the extended phase, mobility edge phase and localized phase, respectively,
		which satisfy the conditions in Eqs.~(\ref{eq:GAM1}) and (\ref{eq:GAM2}).
		\label{fig:M41}}
\end{figure}
\begin{figure}
	\begin{centering}
		\includegraphics[scale=0.48]{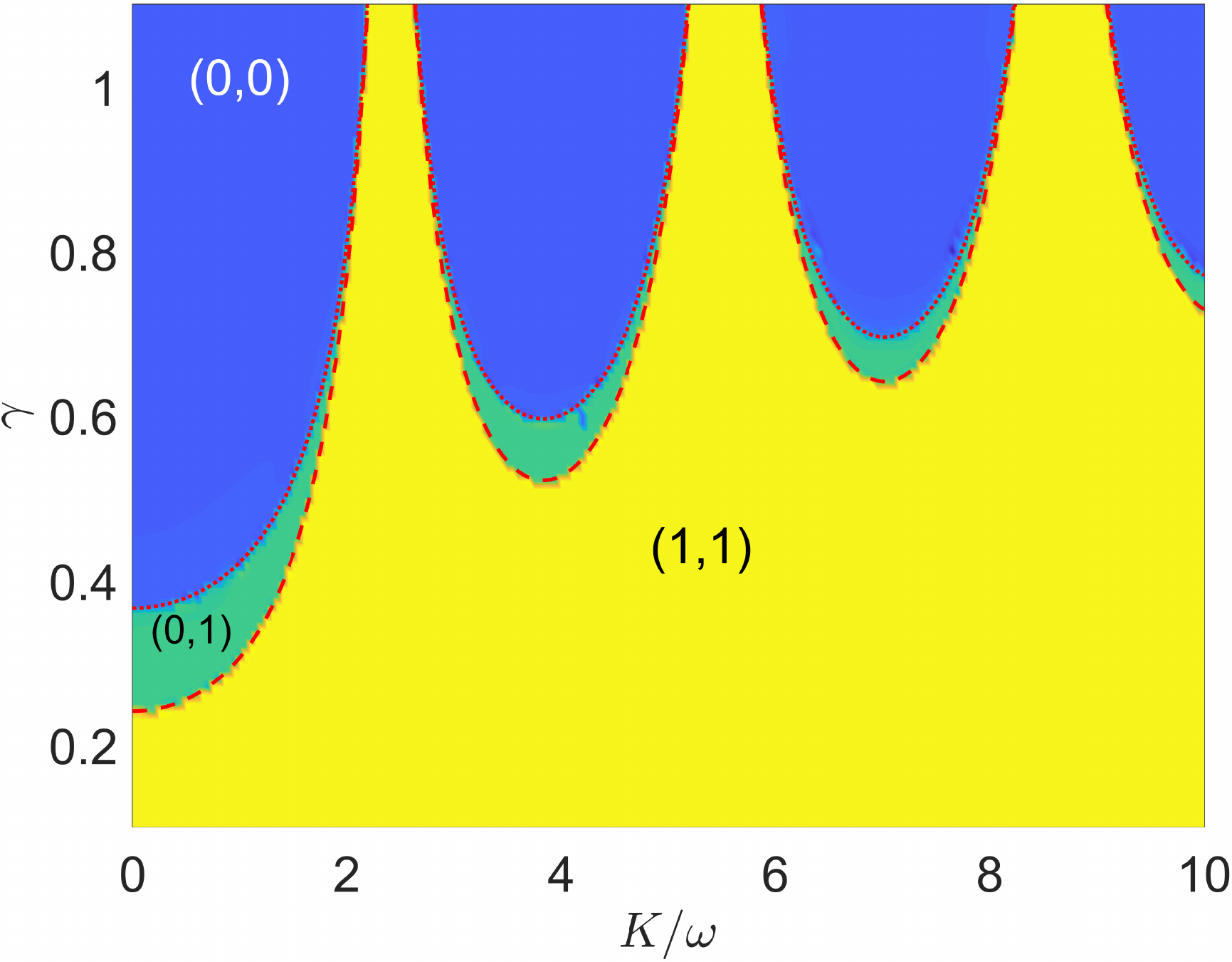}
		\par\end{centering}
	\caption{The topological phase diagram of the non-Hermitian Floquet quasicrystal M4.
		System parameters are chosen as $\alpha=\frac{\sqrt{5}-1}{2}$,
		$J=2$, $V=1$, and the length of lattice is taken as $L=377$.
		Each region with a uniform color corresponds to a phase, 
		whose winding numbers $(w_1,w_2)$ are denoted explicitly therein.
		The red dashed and dotted lines are phase boundaries satisfying
		Eqs.~(\ref{eq:GAM1}) and (\ref{eq:GAM2}), respectively.
		\label{fig:M42}}
\end{figure}

Finally, following the routines of Subsecs. \ref{subsec:M1}\textendash \ref{subsec:M3},
we could construct a pair of spectral winding numbers $(w_{1},w_{2})$
to characterize the topological nature of transitions between different
phases (see Table \ref{tab:ResM4} for their definitions). In
the localized phase, both the base energies $E_{1}=iV$ and $E_{2}=2J{\cal J}_{0}(K/\omega)+iV$
are encircled once by the spectrum of $H_{4{\rm F}}(\theta)$ when
$\theta$ changes over a cycle from zero to $\pi$, and the winding
numbers $(w_{1},w_{2})=(1,1)$. In the mobility edge phase, only $E_{2}=2J{\cal J}_{0}(K/\omega)+iV$
is encircled once by the spectral of $H_{4{\rm F}}(\theta)$ for $\theta\in[0,\pi]$,
and the winding numbers $(w_{1},w_{2})=(0,1)$. In the extended phase,
none of the bases energies $E_{1,2}$ are encircled by the spectral
and we have $(w_{1},w_{2})=(0,0)$. Therefore, these winding numbers
could be employed to distinguish phases with different transport properties
in the M4. For the ease of reference,
we summarize the main results of the M4 in Table~\ref{tab:ResM4}.

To verify our theoretical results, we present numerical calculations
of the IPRs, Lyapunov exponents and winding numbers of the non-Hermitian
Floquet quasicrystal M4 versus $K/\omega$ and $\gamma$ in Figs.~\ref{fig:M41} and \ref{fig:M42}.
We find that states with the minimum of IPRs $\simeq 0$ and Lyapunov exponent $\lambda_{\min}\simeq 0$ start
to emerge when $\gamma$ passes over $\gamma_1$ from below, as shown in Figs.~\ref{fig:M41}(a) and \ref{fig:M41}(c).
This means that some states of the system become delocalized after the first transition at $\gamma=\gamma_1$ (red dashed line).
When $\gamma$ further increases to $\gamma_2$~(red dotted line), a second transition happens and all states become extended with
max(IPR) $\simeq 0$ and $\lambda_{\max}\simeq 0$ after the transition ($\gamma>\gamma_2$), as shown in Figs.~\ref{fig:M41}(b) 
and \ref{fig:M41}(d). The boundaries between the three distinct quasicrystal phases are precisely determined by 
Eqs.~(\ref{eq:GAM1}) and (\ref{eq:GAM2}). Besides, the system could also enter and leave the localized, mobility edge and 
extended phases alternately with the increase of $K/\omega$, and the regime of mobility edge is modified at different values of $K/\omega$.
Therefore, both the localization-delocalization transitions and mobility edges in the non-Hermitian Maryland model M4 could be controlled by the driving field.

In Fig.~\ref{fig:M42}, we compute the winding numbers $(w_1,w_2)$ of M4 following their definitions in Table~\ref{tab:ResM4}. We observe that $(w_1,w_2)=(1,1)$
for the localized phase (below the red dashed line), $(w_1,w_2)=(0,1)$ for the mobility edge phase (between the red dashed and dotted lines),
and $(w_1,w_2)=(0,0)$ for the extended phase (above the red dotted line). These results confirm that the winding numbers $(w_1,w_2)$
could be utilized to distinguish quasicrystal phases of the non-Hermitian Floquet Maryland model with different transport nature, and characterize the
transitions between them. Notably, the winding numbers $(w_1,w_2)$ could also change with the increase of the ratio between the amplitude and frequency of the forcing $K/\omega$, which means that the topological properties of the non-Hermitian Maryland model can be engineered by the driving field.

\subsection{M5: Floquet spectrum, localization transition, mobility edge and topological invariant\label{subsec:M5}}

\begin{table*}
	\begin{centering}
		\begin{tabular}{|c|c|c|c|}
			\hline 
			Phase & Extended & Mobility Edge & Localized\tabularnewline
			\hline 
			\hline 
			Condition & $E_{r}^{\max}<E_{c}$ & $E_{r}^{\min}<E_{c}<E_{r}^{\max}$ & $E_{c}<E_{r}^{\min}$\tabularnewline
			\hline 
			IPR & $\simeq0$ for all states & $>0$ \& $\simeq0$ coexist & $>0$ for all states\tabularnewline
			\hline 
			Mobility edge & \multicolumn{3}{c|}{$E_{c}=|J{\cal J}_{0}(K/\omega)(\eta+1/\eta)|$}\tabularnewline
			\hline
			Quasienergy & Real & \multicolumn{2}{c|}{Complex}\tabularnewline
			\hline  
			Winding number & \multicolumn{3}{c|}{$w=\int_{0}^{2\pi}\frac{d\theta}{2\pi i}\partial_{\theta}\ln\det[H_{5F}(\theta)-E_{c}^{+}]=\begin{cases}
				1 & {\rm Mobility\,\,edge}\\
				0 & {\rm {\rm Otherwise}}
				\end{cases}$}\tabularnewline
			\hline 
		\end{tabular}
		\par\end{centering}
	\caption{Summary of results for the non-Hermitian Floquet quasicrystal M5.
		$J$ is the nearest-neighbor hopping amplitude, $\eta$ controls the
		non-Hermiticity, $K$ is the driving amplitude, $\omega$ is the driving
		frequency, and ${\cal J}_{0}(K/\omega)$ denotes the Bessel function
		of first kind. $H_{5{\rm F}}(\theta)$
		is obtained from $H_{5{\rm F}}$ after setting $V_{n}=V/[1-\eta e^{i(2\pi\alpha n+\theta/L)}]$,
		with $V$ being the amplitude of onsite potential and $L$ the length
		of lattice. $E_{c}^{+}=E_{c}+i0^{+}$ is the base quasienergy with
		respect to which the winding number $w$ is defined.\label{tab:ResM5}}
\end{table*}

\begin{figure}
	\begin{centering}
		\includegraphics[scale=0.46]{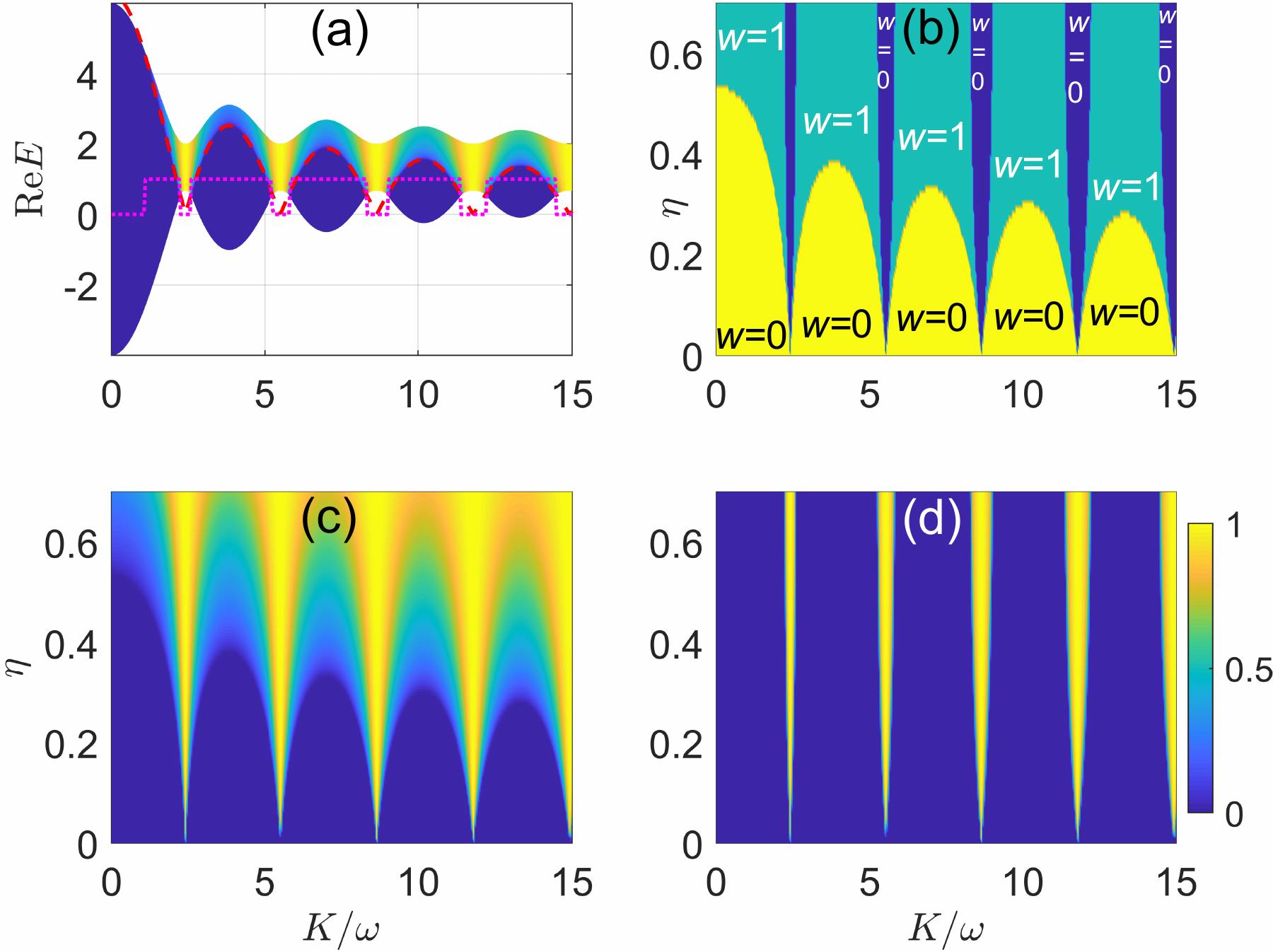}
		\par\end{centering}
	\caption{The real part of quasienergies and IPRs of the non-Hermitian Floquet
		quasicrystal M5. Other system parameters are set as $\alpha=\frac{\sqrt{5}-1}{2}$,
		$J=2.5$, $V=1$, and the length of lattice is taken as $L=610$.
		The color bars in panels (a), (c) and (d) are taken from zero to one.
		In panel (a), the real part of quasienergies ${\rm Re}E$ of all states
		are plotted with respect to $K/\omega$ at $\eta=0.5$, and the color
		of each state represents its IPR. The red dashed and magenta dotted
		lines show the mobility edge $E_{c}$ and winding number $w$ versus
		$K/\omega$. The phase diagram in panel (b) is obtained by evaluating
		${\rm sign}(E_{c}-E_{r}^{\max})+{\rm sign}(E_{c}-E_{r}^{\min})$,
		which is equal to $2$ (yellow regions), $0$ (green regions) and
		$-2$ (blue regions) for the extended, mobility edge and localized
		phases, respectively. The winding number is $w=1$ for the mobility
		edge phase and $w=0$ otherwise. The maximum and minimum of IPRs over
		all states are shown in panels (c) and (d), respectively.
		The mobility edge phase (localized phase) sets in when the
		maximum (minimum) of IPRs starts to deviate from zero.\label{fig:M5}}
\end{figure}

To give another transparent view on how the mobility edge is modified by the driving field, 
we consider again a variant of the driven non-Hermitian AAH model, 
whose Hamiltonian in lattice representation takes the form
\begin{alignat}{1}
\hat{H}_{5}(t)= & \sum_{n}J\left(\hat{c}_{n}^{\dagger}\hat{c}_{n+1}+{\rm h.c.}\right)\nonumber \\
+ & \sum_{n}\left[V/(1-\eta e^{i2\pi\alpha n})-nK\cos(\omega t)\right]\hat{c}_{n}^{\dagger}\hat{c}_{n}.\label{eq:H5t}
\end{alignat}
Here the parameter $\eta$ controls the amount of non-Hermiticity,
and the system is quasiperiodic when $\alpha$ is irrational.
In the high-frequency limit, we obtain the Floquet effective Hamiltonian of this M5 as
\begin{alignat}{1}
\hat{H}_{5{\rm F}}= & \sum_{n}J{\cal J}_{0}\left(K/\omega\right)\left(\hat{c}_{n}^{\dagger}\hat{c}_{n+1}+{\rm h.c.}\right)\nonumber \\
+ & \sum_{n}V/(1-\eta e^{i2\pi\alpha n})\hat{c}_{n}^{\dagger}\hat{c}_{n},\label{eq:H5F}
\end{alignat}
where the hopping amplitude is controlled by the ratio between the amplitude
and frequency of the driving force. Acting $\hat{H}_{5{\rm F}}$ on the state 
$|\psi\rangle=\sum_{n}\psi_{n}\hat{c}_{n}^{\dagger}|0\rangle$, we find 
the eigenvalue equation
\begin{equation}
J{\cal J}_{0}\left(\frac{K}{\omega}\right)(\psi_{n+1}+\psi_{n-1})+\frac{V}{1-\eta e^{i2\pi\alpha n}}\psi_{n}=E\psi_{n},\label{eq:H5E}
\end{equation}
whose solutions yield the quasienergy $E$ and Floquet eigenstates of the system. 
Note that since the onsite potential amplitude $V_n=V^*_{-n}$, the system Hamiltonian also
possesses the ${\cal PT}$-symmetry and its spectrum could be real 
in certain parameter regime.

Following the method outlined in Ref.~\cite{LonghiQC3}, the non-Hermitian Floquet quasicrystal M5
is found to have a quasienergy-dependent mobility edge at $E=E_c$, where
\begin{equation}
E_c=|J{\cal J}_0(K/\omega)(\eta+1/\eta)|.\label{eq:M5ME}
\end{equation}
Let us denote $E_r^{\max}$ and $E_r^{\min}$ as the maximum and minimum
of the real parts of quasienergies over all states at a fixed set
of system parameters. Then if $E_r^{\max}<E_c$, all eigenstates of
the system are delocalized and the system sits in an extended phase.
When $E_r^{\min}<E_c<E_r^{\max}$, we find the states with ${\rm Re}E<E_c$~(${\rm Re}E>E_c$)
to be extended~(localized). A mobility edge thus emerges at $E=E_c$, and 
the system resides in the mobility edge phase under the condition $E_r^{\min}<E_c<E_r^{\max}$.
When $E_c<E_r^{\min}$, the system enters a phase in which all eigenstates
are localized, and the mobility edge vanishes. By tuning the driving field
parameters $K$ and $\omega$, the system could undergo alternate transitions
among the extended, mobility edge and localized phases. The Floquet engineering
scheme can again be utilized to dynamically control the mobility edges and transport nature of the
non-Hermitian Floquet quasicrystal M5. We summarize our key results about
this model in Table~\ref{tab:ResM5}.

To demonstrate our theoretical findings, we compute the spectrum and IPRs of M5, as presented in
Fig.~\ref{fig:M5}. Note that in Fig.~\ref{fig:M5}(b), we evaluated a winding number $w$ of the 
quasienergy with respect to the mobility edge $E_c$~(see Table~\ref{tab:ResM5} for its definition),
and find that $w=1$ for the mobility edge phase and $w=0$ otherwise. This winding number
can thus be used as a topological order parameter to distinguish the mobility edge phase
and other phases of the system, and to signify the transitions bewteen them.
The numerical results shown in Fig.~\ref{fig:M5} are all consistent with our theoretical
predictions about the non-Hermitian Floquet quasicrystal M5.
Combining these observations with the results obtained in Subsecs.~\ref{subsec:M1}--\ref{subsec:M4},
we conclude that the Floquet engineering strategy could indeed be exploited
as a useful means to control and modulate the spectrum, topological
and transport properties in a broad class of non-Hermitian quasicrystals.

\section{Summary and discussion\label{sec:Sum}}

In this work, we apply Floquet driving fields to engineer ${\cal PT}$-breaking,
localization-to-delocalization, topological transitions and mobility edges in non-Hermitian
quasicrystals. Following the well-known scheme of dynamical localization,
we utilize high-frequency shaking forces to control the hopping amplitudes
of non-Hermitian quasiperiodic lattices, and observing alternate transitions
between extended, localized and mobility edge phases with the change of driving field
parameters. We implement our scheme in five prototypical models of
non-Hermitian quasicrystal, and obtain the conditions of their driving-assisted
${\cal PT}$-transitions, localization-delocalization transitions and mobility edges. 
We further introduced winding numbers of the quasienergy
to distinguish non-Hermitian Floquet quasicrystals with different
spectrum and localization features, and establishing the topological phase diagrams
of these intriguing phases. Our results thus extend the study of non-Hermitian
quasicrystals to periodically driven systems, and further uncover
the usefulness of the Floquet engineering approach in the design and
control of non-Hermitian quasicrystals with distinctive topological
and transport properties.

In theory, the critical exponent of localization length~(inverse Lyapunov exponent) could characterize the critical nature of delocalization-localization transitions~\cite{ChayesIneq}. For the original AAH model (see Eq.~(\ref{eq:HAAH})) we have $\lambda=\ln(V/2J)$, and $\lambda^{-1}\propto|V-V_c|^{-1}$ around the critical point $V_c=2J$ of the transition. The critical exponent is thus $\nu=1$. Referring to the tables \ref{tab:ResM1}--\ref{tab:ResM3}, we find that close to the critical points of models M1--M3, the localization length scales as $\lambda^{-1}\propto|V-V_c|^{-1}$, $\lambda^{-1}\propto|\gamma-\gamma_c|^{-1}$ and $\lambda^{-1}\propto|(K/\omega)-(K/\omega)_c|^{-1}$. Therefore, the critical exponent of these models is also $\nu=1$. In Ref.~\cite{ChayesIneq}, it was found that for 1D systems with \textit{independent} bond or site disorder, the critical exponent satisfies the Chayes' inequality $\nu\geq2$. For 1D quasicrystals, since the disorder introduced by the quasiperiodic potential can be viewed as \textit{long-range correlated}, the Chayes' inequality may not hold. Our results thus provide examples for the violation of the Chayes' inequality in 1D non-Hermitian Floquet quasicrystals.

Besides discrete lattices, localization transitions in quasicrystals have also been explored in continuum models~\cite{ContinuumQC1,ContinuumQC2,ContinuumQC3}. A generic observation for continuum models is that mobility edges will emerge beyond the tight-binding limit. Since we focused on the high-frequency region of the driving field in our study, the resulting systems are described by effective Floquet Hamiltonians with modified hopping amplitudes. Therefore, we expect that with long-range hoppings or in continuum versions of the models M1--M3, mobility edges may also appear, and the physical mechanism is similar to that discovered in Refs.~\cite{ContinuumQC1,ContinuumQC2,ContinuumQC3}. In the meantime, the transition of the spectrum from real to complex in continuum versions of M1--M3 would be related to the transition from an extended or a localized phase to a mobility edge phase. Due to the possible presence of mobility edges, the energy-independent Lyapunov exponents shown in tables \ref{tab:ResM1}--\ref{tab:ResM3} may not be valid for continuum models, and should be reconsidered case by case.

In future work, it would be interesting to apply our scheme to the
engineering of non-Hermitian quasicrystals with other types of mobility edges, long-range hoppings, 
many-body effects and in higher spatial dimensions.
In the present study, the experimentally established high-frequency approach~\cite{DLRev2} is applied to control the tunneling amplitudes, and we expect it to be valid under the condition listed below Eq.~(\ref{eq:HR2}). Interestingly, our initial results~(in preparation) suggest that driving fields could induce transitions among extended, localized and mobility edge phases at generic frequencies. The results presented in Figs.~\ref{fig:M1}--\ref{fig:M5} indicate that this general feature has already been captured qualitatively under the high-frequency approximation.
Under moderate driving frequencies, quasicrystalline phases that are unique to non-Hermitian Floquet systems may also emerge, which certainly deserve more thorough explorations.

\begin{acknowledgments}
L.Z. is supported by the National Natural Science Foundation of China
(Grant No. 11905211), the China Postdoctoral Science Foundation (Grant
No. 2019M662444), the Fundamental Research Funds for the Central Universities
(Grant No. 841912009), the Young Talents Project at Ocean University
of China (Grant No. 861801013196), and the Applied Research Project
of Postdoctoral Fellows in Qingdao (Grant No. 861905040009).
\end{acknowledgments}

\end{document}